\documentclass[12pt]{article}
\usepackage{amsfonts,amsmath,amsthm}
\usepackage{graphicx,psfrag,epsf}
\usepackage{enumerate}
\usepackage{dsfont}
\usepackage[margin=.7in]{geometry}

\usepackage [english]{babel}
\usepackage [autostyle, english = american]{csquotes}
\MakeOuterQuote{"}

\RequirePackage[colorlinks,citecolor=blue,urlcolor=blue]{hyperref}
\RequirePackage[numbers]{natbib}

\usepackage{fancyhdr}
\newtheorem{thm}{Theorem}
\newtheorem{cor}{Corollary}
\newtheorem{lemma}{Lemma}
\newtheorem{proposition}{Proposition}

\numberwithin{equation}{section}
\numberwithin{thm}{section}
\numberwithin{cor}{section}
\numberwithin{lemma}{section}
\numberwithin{proposition}{section}
\theoremstyle{plain}

\begin{document}

{
  \title{\bf Dynamic tail inference \\ with log-Laplace volatility\footnote{Preprint}}
  \author{Gordon V. Chavez\footnote{gchavez@novocure.com}}

\date{}
  \maketitle
} 

\bigskip
\begin{abstract}

We propose a family of models that enable predictive estimation of time-varying extreme event probabilities in heavy-tailed and nonlinearly dependent time series. The models are a white noise process with conditionally log-Laplace stochastic volatility. In contrast to other, similar stochastic volatility formalisms, this process has analytic expressions for its conditional probabilistic structure that enable straightforward estimation of dynamically changing extreme event probabilities. The process and volatility are conditionally Pareto-tailed, with tail exponent given by the reciprocal of the log-volatility's mean absolute innovation. This formalism can accommodate a wide variety of nonlinear dependence, as well as conditional power law-tail behavior ranging from weakly non-Gaussian to Cauchy-like tails. We provide a computationally straightforward estimation procedure that uses an asymptotic approximation of the process' dynamic large deviation probabilities. We demonstrate the estimator's utility with a simulation study. We then show the method's predictive capabilities on a simulated nonlinear time series where the volatility is driven by the chaotic Lorenz system. Lastly we provide an empirical application, which shows that this simple modeling method can be effectively used for dynamic and predictive tail inference in financial time series.

\end{abstract}



\newpage
\section{Introduction}
\label{sec:intro}

Financial time series data is well-known to exhibit nonlinear dependence and "fat tails". Such time series often display trends of increasing or decreasing volatility, along with a propensity for extreme fluctuations that is far greater than what would be predicted from a Gaussian or other distribution with finite polynomial moments. The latter observation was probably most famously addressed with the early Pareto-tailed and stable models for financial time series proposed by Mandelbrot (1963) and Fama (1968), while the most well-known early approaches to the nonlinear dependence problem were given by Engle (1982) and Bollerslev (1986) with original and generalized autoregressive conditional heteroskedasticity (ARCH, GARCH) models. Since then a great deal of research has been dedicated to extreme event probability estimation and nonlinear time series modeling for financial applications (e.g., Embrechts et al. 2011 and Terasvirta et al. 2010). 

Estimation of extreme event probabilities is a very important problem for risk and portfolio management. Many estimators for the tail exponent of a marginal distribution have been proposed, e.g., by Hill (1975), Pickands (1975), and deHaan and Resnick (1980), however, these estimators are very sensitive to dependence in the data (Kearns and Pagan 1997, Diebold et al. 2000). This makes them often ill-suited for application to many strongly dependent time series of interest for financial modeling, i.e., the squares and moduli of financial log-returns (Embrechts et al. 2011 p. 270, 406). Relatedly, these estimators along with much of extreme value theory (EVT) are designed for inference of stationary, rather than time-varying, tail behavior. Gardes and Girard (2008) and Gardes and Stupfler (2014) have given nonparametric estimators for time-varying tail exponents, while Kelly (2014) has given a parametric approach to dynamic power law estimation in financial time series. The parametric modeling method we propose here, however, does not assume a time-varying tail exponent. Our approach is hence closer to McNeil and Frey's (2000) combination of stationary EVT with GARCH modeling. However, we use a novel, stochastic volatility approach to enable dynamic tail inference. 

A canonical form of stochastic volatility model, first proposed by Taylor (1982, 1986), is given by
\begin{equation}
\varepsilon_{t}=\sigma_{t} z_{t}, \label{process}
\end{equation}
where $z_{t}$ is an \textit{i.i.d.} process with a mean of 0 and a variance of 1, and $\sigma_{t}$ is a non-negative process defined by 
\begin{equation}
\sigma_{t}=\exp\left(H_{t}\right), \label{volatility}
\end{equation}
where $H_{t}$ is a Gaussian process with mean $\mu_{H}$ and variance $\sigma_{H}^{2}<\infty$. An important example was the AR(1) model $H_{t}=\mu_{H}+\beta \left(H_{t-1}-\mu_{H} \right)+h_{t}$, where $\beta \in \mathbb{R}$ and $h_{t} \sim \mathcal{N} \left(0, \sigma_{h}^{2} \right)$ is \textit{i.i.d.} The kind of model in (\ref{process}) and (\ref{volatility}), where $H_{t}$ is a variety of Gaussian processes, has been extensively applied and studied, e.g., to exchange rate modeling by Harvey et al. (1994) and extended to long-memory Gaussian $H_{t}$ by Breidt et al. (1998). However, the stochastic volatility $\sigma_{t}$ in (\ref{volatility}) follows a log-normal distribution, which has finite polynomial moments (see Johnson et al. 1994). As a consequence, (\ref{process}) also has bounded moments in the usual case of Gaussian $z_{t}$. This can be problematic for modeling time series with power law tails and divergent higher-order polynomial moments. Practically, such models will underestimate the probabilities of extreme events. To remedy this, $z_{t}$ has often been chosen to follow the heavier-tailed Student's $t$-distribution, e.g., in Harvey et al. (1994), Liesenfeld and Jung (2000), and Chib et al. (2002). However, with a Gaussian or Student's $t$ choice for $z_{t}$ and Gaussian $H_{t}$, the model (\ref{process}) does not have convenient analytic expressions for its conditional probabilistic structure. This makes estimation of model parameters as well as outcome probabilities difficult, often requiring the use of Bayesian and Monte Carlo methods such as those described in Jacquier et al. (1994), Kim et al. (1998), Sandmann and Koopman (1998), or Chib et al. (2002). Reliable estimation of similar, conditionally Student's t-distributed, ARCH-related models requires similar numerical procedures (see, e.g., Mousazadeh and Karimi 2007, Ardia 2008, Ardia and Hoogerheide 2010).

In this paper we propose a stochastic volatility formalism that enables straightforward and effective estimation of time-varying extreme event probabilities in time series with power law tail behavior. The models we present have the form (\ref{process})-(\ref{volatility}), with $z_{t}\sim \mathcal{N}(0,1)$. However, instead of defining (\ref{volatility})'s $H_{t}$ as Gaussian, we make $H_{t}$ conditionally \textit{Laplace}-distributed, defining $H_{t}$ as
\begin{equation}
H_{t}=E\left \{H_{t}|\mathcal{F}_{t-1}\right\}+h_{t}, \label{entropy}
\end{equation}
where $\mathcal{F}_{t-1}$ is the filtration up to time $t-1$ and $h_{t}$ is \textit{i.i.d.} Laplace-distributed with density
\begin{equation}
p_{h}\left(h_{t}\right)=\frac{1}{2\Delta} \exp\left(-\frac{\left|h_{t}\right|}{\Delta}\right), \label{laplace}
\end{equation}
where
$$\Delta=E\left\{\left|h_{t}\right|\right\}.$$ 
This simple but important adjustment endows (\ref{process})-(\ref{volatility})'s $\sigma_{t}$ and $\varepsilon_{t}$ with power law-tailed conditional probability distributions, for which there are natural and convenient analytic expressions. These conditional distributions give the result
\begin{equation}
P\left\{\left|\varepsilon_{t}\right| \geq \Lambda |\mathcal{F}_{t-1} \right\} \sim f(\Delta) \exp\left(\frac{E\left \{H_{t}|\mathcal{F}_{t-1}\right\}}{\Delta}\right)\Lambda^{-1/\Delta}
\label{tailresult}
\end{equation}
as $\Lambda \rightarrow \infty$. Hence $H_{t}$'s mean absolute innovation $\Delta$ specifies the tail exponent, which allows (\ref{process})-(\ref{laplace}) to flexibly define processes ranging from only mildly non-Gaussian with $\Delta \approx 0$ to processes with Cauchy-like tails at $\Delta=1$. The process $\varepsilon_{t}$'s tail probabilities are also strongly and explicitly dependent on the process $H_{t}$, which enables their dynamic estimation. We give a simple, probabilistic method-of-moments estimation procedure for the tail exponent, which takes advantage of the result (\ref{tailresult}) for $\varepsilon_{t}$'s tail probabilities. We present a simulation study to show the effectiveness of this estimator. We next demonstrate the predictive capabilities of this methodology with a simulated nonlinear time series, where the volatility is driven by the chaotic Lorenz system. We then give an empirical application to S\&P 500 Index data, which shows that this modeling method can be used for dynamic and predictive estimation of volatility and extreme event probabilities in heavy-tailed financial time series. We give some concluding remarks, an appendix with proofs of the main results, and a second appendix, which shows that very similar results, including (\ref{tailresult}), can also be derived for the case when $z_{t}$ is Laplace-distributed.

\section{Model}
\label{sec: model}
We will denote $H_{t}$'s conditional expectation as $\bar{H}_{t}=E\left \{H_{t}|\mathcal{F}_{t-1}\right\}$. We first give expressions for $\sigma_{t}$'s conditional probability density and $\varepsilon_{t}$'s conditional polynomial moments.

\begin{lemma}
Given (\ref{entropy})-(\ref{laplace}), $\sigma_{t}$ in (\ref{volatility}) is conditionally distributed according to the log-Laplace probability density function
\begin{equation}
p_{\sigma}(\sigma_{t}|\mathcal{F}_{t-1})=\left\{\begin{matrix} \frac{1}{2\Delta}\exp\left(-\frac{\bar{H}_{t}}{\Delta}\right) \sigma_{t}^{1/\Delta-1};\hspace{.3cm} 0< \sigma_{t} <\exp\left(\bar{H}_{t}\right)

\\ 
\frac{1}{2 \Delta} \exp\left(\frac{\bar{H}_{t}}{\Delta}\right) \sigma_{t}^{-1/\Delta-1};\hspace{.3cm} \sigma_{t} \geq \exp\left(\bar{H}_{t}\right)

\end{matrix}\right. \label{loglaplace}
\end{equation} \label{lem2.1}
\end{lemma}

\begin{figure}
\includegraphics[width=\linewidth]{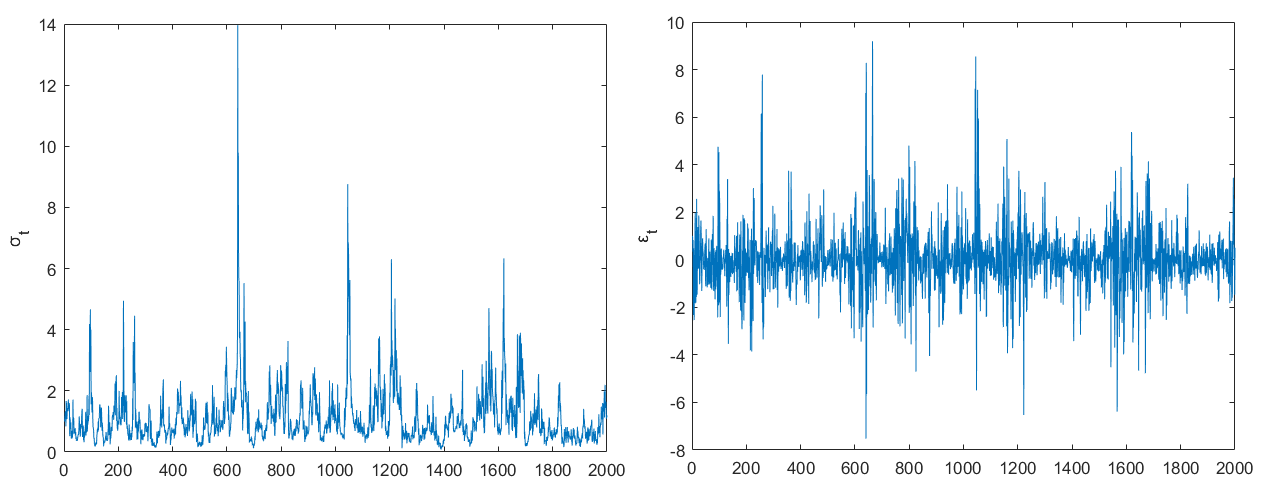}
\caption{A typical realization of the stochastic volatility $\sigma_{t}$ defined in (\ref{volatility})-(\ref{laplace}) and (\ref{process})'s corresponding process $\varepsilon_{t}$ with $z_{t}\sim \mathcal{N}(0,1)$, $H_{t}=.5H_{t-1}+.4H_{t-2}+h_{t}$, and $\Delta=1/4$.}
\label{fig: 0}
\end{figure}

\begin{figure}
\includegraphics[width=\linewidth]{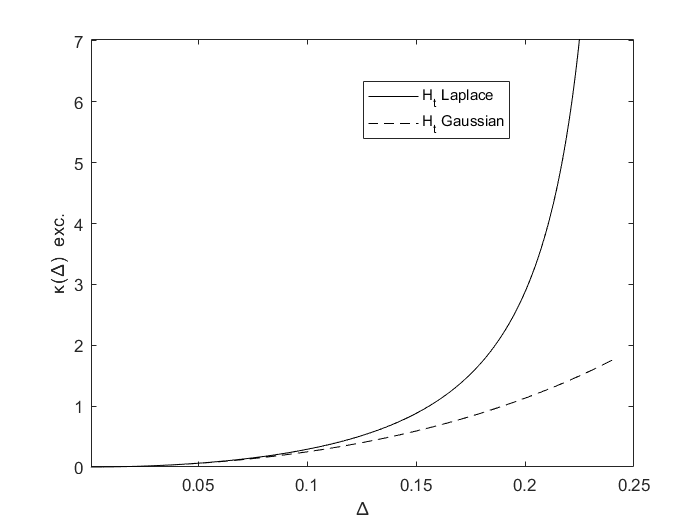}
\caption{A graph of (\ref{kurtosis})'s conditional excess kurtosis $\kappa^{\text{exc.}}(\Delta)=\kappa(\Delta)-3$ for Laplace $H_{t}$ (Solid) along with the corresponding result for Gaussian $H_{t}$ (Dashed).}
\label{fig: 1}
\end{figure}

\newpage

\begin{cor}
Given (\ref{process}) and (\ref{loglaplace}) with $z_{t}\sim \mathcal{N}(0,1)$, for all even $n\geq 2$, if $\Delta \geq 1/n$, then $E\left \{\varepsilon_{t}^{n}|\mathcal{F}_{t-1}\right \}=\infty$, while if $\Delta < 1/n$,
\begin{equation}
E\left \{\varepsilon_{t}^{n}|\mathcal{F}_{t-1}\right\}=\exp\left(n\bar{H}_{t}\right)\frac{(n-1)!!}{1-n^{2}\Delta^{2}}. \label{processmoment}
\end{equation} \label{cor2.1}
\end{cor}

\begin{figure}
\includegraphics[width=\linewidth]{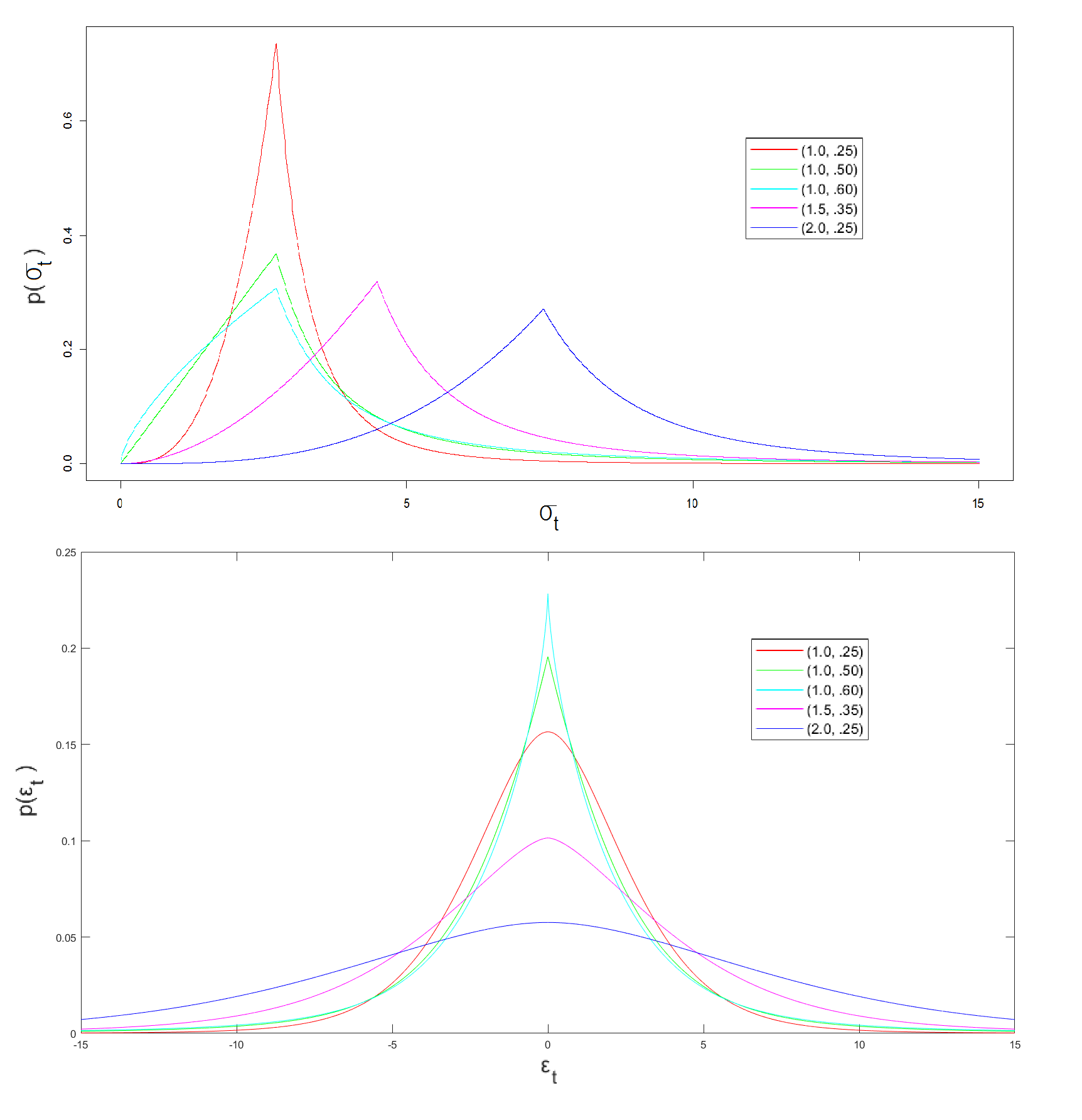}
\caption{Graphs of the probability density functions (\ref{loglaplace}) (Top) and (\ref{bigprob}) (Bottom) with $\left(\bar{H}_{t},\Delta\right)$ equal to (1, .25) in Red, (1, .50) in Green, (1, .60) in Cyan, (1.5, .35) in Magenta, and (2, .25) in Blue.}
\label{fig: 2}
\end{figure}

Several graphs of (\ref{loglaplace}) are given in Fig. \ref{fig: 2}. It is clear from Corollary \ref{cor2.1} that the conditional kurtosis $\kappa_{t}=E\left\{\varepsilon_{t}^{4}|\mathcal{F}_{t-1}\right\}/\left(E\left\{\varepsilon_{t}^{2}|\mathcal{F}_{t-1}\right\}\right)^{2}$
diverges or is not well-defined for $\Delta \geq 1/4$. However, when (\ref{processmoment}) is used to calculate $\kappa_{t}$ for $\Delta < 1/4$, it can be easily shown that the conditional kurtosis has the time-independent definition
\begin{equation}
\kappa(\Delta)=3\frac{\left(1-4\Delta^{2}\right)^{2}}{1-16\Delta^{2}}, \label{kurtosis}
\end{equation}
which is minimized at $\kappa(0)=3$, the Gaussian kurtosis. The divergent algebraic structure of (\ref{kurtosis}) contrasts with the kurtosis' exponential structure for Gaussian $H_{t}$, which would be written as $\kappa_{t}=3\exp\left(8\Delta^{2}\right)$. The \textit{excess kurtosis} $\kappa^{\text{exc.}}(\Delta)=\kappa(\Delta)-3$ is graphed in Fig. \ref{fig: 1} for both Laplace and Gaussian $H_{t}$. Now that we have given $\varepsilon_{t}$'s conditional moment structure, we will give its conditional probability density function.

\begin{thm}
Given (\ref{loglaplace}) with $z_{t}\sim \mathcal{N}(0,1)$, (\ref{process})'s $\varepsilon_{t}$ is conditionally distributed according to the probability density function
\begin{eqnarray}
p_{\varepsilon}\left(\varepsilon_{t}|\mathcal{F}_{t-1}\right)=\frac{1}{4 \sqrt{\pi} \Delta} \left(\sqrt{2}e^{\bar{H}_{t}}\right)^{-1/\Delta}\Gamma \left(\frac{1-1/\Delta}{2},\frac{\varepsilon_{t}^{2}}{2e^{2\bar{H}_{t}}}\right)\left|\varepsilon_{t} \right|^{1/\Delta-1} \label{bigprob} \\ \nonumber +\frac{1}{4 \sqrt{\pi} \Delta}\left(\sqrt{2}e^{\bar{H}_{t}}\right)^{1/\Delta} \gamma \left(\frac{1+1/\Delta}{2},\frac{\varepsilon_{t}^{2}}{2e^{2\bar{H}_{t}}} \right)\left|\varepsilon_{t} \right|^{-1/\Delta-1} 
\end{eqnarray}
where 
\begin{equation}
\Gamma(a,b)=\int_{b}^{\infty}x^{a-1}e^{-x}dx \label{upgamma}
\end{equation}
is the \textit{upper incomplete gamma function} and 
\begin{equation}
\gamma(a,b)=\int_{0}^{b} x^{a-1}e^{-x}dx \label{lowgamma}
\end{equation}
is the \textit{lower incomplete gamma function}. \label{thm2.1}
\end{thm}
Several graphs of (\ref{bigprob}) are given in Fig. \ref{fig: 2}. Note that with larger $\Delta$ and smaller $\bar{H}_{t}$ the densities are more sharply peaked around the origin. This can also be seen from the following result.

\begin{cor}
For $\Delta<1$, as $\left|\varepsilon_{t}\right|\rightarrow 0$, 
\begin{equation}
p_{\varepsilon}\left(\varepsilon_{t}|\mathcal{F}_{t-1}\right) \sim \frac{1}{\sqrt{2\pi}e^{\bar{H}_{t}}}\frac{1}{1-\Delta^{2}}.
\label{near0res}
\end{equation}
\label{cor2.2}
\end{cor}

The above result characterizes $\varepsilon_{t}$'s near-mean behavior. Next we give an analytic expression for $\left|\varepsilon_{t}\right|$'s exceedance probabilities, before characterizing $\varepsilon_{t}$'s tail behavior. We first make the notation 
\begin{equation}
\widetilde{\Lambda}_{t}=\frac{\Lambda}{\sqrt{2}e^{\bar{H}_{t}}}, \label{lamsub}
\end{equation}
which will make several of the next results more symbolically concise. 

\begin{thm}
By (\ref{bigprob})-(\ref{lowgamma}), 
\begin{equation}
P\left \{\left|\varepsilon_{t}\right| \geq \Lambda|\mathcal{F}_{t-1}\right \}=\frac{1}{2\sqrt{\pi}}\left(\gamma\left(\frac{1+1/\Delta}{2},\widetilde{\Lambda}_{t}^{2}\right)\widetilde{\Lambda}_{t}^{-1/\Delta}-\Gamma\left(\frac{1-1/\Delta}{2},\widetilde{\Lambda}_{t}^{2}\right)\widetilde{\Lambda}_{t}^{1/\Delta}\right)+\textup{erfc}\left(\widetilde{\Lambda}_{t}\right), \label{datclosedform}
\end{equation}
where 
\begin{equation}
\textup{erfc}(x)=1-\textup{erf}(x)=1-\frac{2}{\sqrt{\pi}}\int_{0}^{x}e^{-u^{2}}du=\frac{2}{\sqrt{\pi}}\int_{x}^{\infty}e^{-u^{2}}du \label{comperf}
\end{equation}
is the complementary error function.
\label{closedformthm}
\end{thm}

Note from the definition (\ref{lowgamma}) that as $\widetilde{\Lambda}_{t} \rightarrow \infty$, (\ref{datclosedform})'s first term will tend to a power law multiplied by the gamma function and other constants. By (\ref{upgamma}) and (\ref{comperf}) however, the second and third term of (\ref{datclosedform}) will asymptotically become much smaller. This leads to the next result, which will be useful for straightforwardly approximating $\varepsilon_{t}$'s dynamic extreme event probabilities.

\begin{thm}
By (\ref{bigprob}),
\begin{equation}
P\left \{\left|\varepsilon_{t}\right| \geq \Lambda|\mathcal{F}_{t-1}\right \}=\frac{1}{2\sqrt{\pi}} \Gamma \left(\frac{1+1/\Delta}{2}\right)\widetilde{\Lambda}_{t}^{-1/\Delta}+O\left(\widetilde{\Lambda}_{t}^{-5}e^{-\widetilde{\Lambda}_{t}^{2}}\right) \label{bigtail} 
\end{equation}
as $\widetilde{\Lambda}_{t}\rightarrow \infty$.
\label{cor2.3}
\end{thm}

This result reduces to the asymptotic power law in (\ref{tailresult}). We will make extensive use of Theorem \ref{cor2.3} in our estimation procedure described in the next section. The result arises from asymptotic expansions of (\ref{bigprob}) and its integral, which are given in (\ref{ukas})-(\ref{pdf}) and (\ref{datseries}) respectively. 
We note from integrating (\ref{loglaplace}) with $\Lambda \geq \exp\left(\bar{H}_{t}\right)$ that 
\begin{equation}
P\left \{\sigma_{t} \geq \Lambda | \mathcal{F}_{t-1} \right \}=\frac{1}{2} \exp\left(\frac{\bar{H}_{t}}{\Delta}\right) \Lambda^{-1/\Delta}. \label{voltail}
\end{equation}
Comparison of (\ref{voltail}) with (\ref{bigtail}) after re-substituting for (\ref{lamsub}) gives 
\begin{equation}
 P\left \{\left|\varepsilon_{t}\right| \geq \Lambda|\mathcal{F}_{t-1}\right \}\sim\frac{\sqrt{2}^{1/\Delta}}{\sqrt{\pi}}\Gamma\left(\frac{1+1/\Delta}{2}\right) P\left \{\sigma_{t} \geq \Lambda|\mathcal{F}_{t-1} \right \}. \label{bigtailvol}
\end{equation}
This result shows that, asymptotically, $\varepsilon_{t}$'s large deviation probabilities are simply proportional to $\sigma_{t}$'s corresponding probabilities through a $\Delta$-dependent factor, and hence the probabilities' dependence on $\bar{H}_{t}$ is given straightforwardly by $\exp\left(\bar{H}_{t}/\Delta\right)$. We make the additional note that $\Gamma(3/2)=\sqrt{\pi}/2$ and $\Gamma(1/2)=\sqrt{\pi}$, therefore (\ref{bigtailvol}) reduces to simply 
$$P\left \{\left|\varepsilon_{t}\right| \geq \Lambda| \mathcal{F}_{t-1}\right \}\sim  P\left \{\sigma_{t} \geq \Lambda|\mathcal{F}_{t-1} \right \}$$
for $\Delta=1/2$ and as $\Delta \rightarrow \infty$. 

It can be seen in (\ref{bigtail}) and (\ref{datseries}) that more complex dependence on $\bar{H}_{t}$ and $\Delta$ is captured by higher-order terms with comparatively small contributions at high $\Lambda$ relative to $e^{\bar{H}_{t}}$.  It is informative to write out the first few terms of (\ref{datseries})'s asymptotic expansion with (\ref{lamsub}), giving 
\begin{eqnarray}
P\left\{\left|\varepsilon_{t}\right| \geq \Lambda|\mathcal{F}_{t-1}\right\}=\frac{1}{2\sqrt{\pi}} \Gamma \left(\frac{1+1/\Delta}{2}\right) \widetilde{\Lambda}_{t}^{-1/\Delta}-\frac{1}{4\sqrt{\pi}\Delta^{2}}\Gamma\left(-\frac{3}{2},\widetilde{\Lambda}_{t}^{2}\right) \label{datserieswrittenout}\\ \nonumber +\frac{1}{2\sqrt{\pi}\Delta^{2}}\Gamma\left(-\frac{5}{2},\widetilde{\Lambda}_{t}^{2}\right)+O\left(\widetilde{\Lambda}_{t}^{-9}e^{-\widetilde{\Lambda}_{t}^{2}}\right).
\end{eqnarray}
Note that the first order correction is negative, indicating that the approximation (\ref{bigtail}) has a small positive bias at large $\widetilde{\Lambda}_{t}$. Also note that the correction terms are inversely proportional to $\Delta^{2}$, meaning that (\ref{bigtail})'s approximation is more accurate for heavier-tailed processes with larger $\Delta$. Now that we have presented the model's polynomial moments and probabilistic structure, we will describe an estimation procedure using some of the above results.

\section{Estimation}
\label{sec: estimation}

We begin by considering the conditional expectation of the unobservable quantity $H_{t}$, given the value of the observable $\log\left|\varepsilon_{t}\right|$.

\begin{proposition}
Given (\ref{process}) and (\ref{volatility}),
\begin{equation}
E\left\{H_{t}| \log \left|\varepsilon_{t}\right|\right\}=\log \left|\varepsilon_{t}\right|+\frac{\log 2+\gamma}{2} \approx \log \left|\varepsilon_{t}\right|+.6352, \label{prehest}
\end{equation}
where 
$$\gamma=\lim_{n \rightarrow \infty} \left(\sum_{k=1}^{n} \frac{1}{k}-\log n \right) \approx .5772$$
is the Euler-Mascheroni constant. \label{prop3.1}
\end{proposition}

By Proposition \ref{prop3.1}, an unbiased estimator for $H_{t}$ is given by
\begin{equation}
\widehat{H}_{t}=\log \left|\varepsilon_{t}\right|+\frac{\log 2+\gamma}{2}\approx \log \left|\varepsilon_{t}\right|+.6352. \label{hest}
\end{equation}
An arbitrary regression model $m(.)$ for estimation of $E\left\{H_{t}|\mathcal{F}_{t-1}\right\}$ may then be trained using the $\widehat{H}_{t}$'s and any other relevant variables $\vec{X}_{t-1},...,\vec{X}_{t-q}$, giving
\begin{equation}
\widehat{\bar{H}}_{t}=\widehat{E}\left\{H_{t}|\mathcal{F}_{t-1}\right\}=m\left(\widehat{H}_{t-1},..., \widehat{H}_{t-p}, \vec{X}_{t-1},...,\vec{X}_{t-q},...\right). \label{regmodel}
\end{equation}
We then estimate $\Delta$ by finding an approximate solution to the empirical, probabilistic moment constraint 
\begin{equation}
\frac{1}{T}\sum_{t=1}^{T}\mathds{1}\left(\left|\varepsilon_{t}\right| \geq \Lambda\right)=\frac{1}{T}\sum_{t=1}^{T}P\left \{\left|\varepsilon_{t}\right| \geq \Lambda|\widehat{\bar{H}}_{t},\widehat{\Delta}\right \},
\label{probconstraint}
\end{equation}
The constraint (\ref{probconstraint}) is justified by the law of total expectation, which requires that 
$E\left\{\mathds{1}\left(\left|\varepsilon_{t}\right| \geq \Lambda\right)\right\}=E\left\{E\left\{\mathds{1}\left(\left|\varepsilon_{t}\right| \geq \Lambda\right)|\mathcal{F}_{t-1} \right\}\right\}$. This estimation method requires a result for the probability estimate on (\ref{probconstraint})'s right-hand side. Using (\ref{datclosedform})'s full result is less computationally straightforward because it is defined in terms of the highly nonstandard incomplete gamma functions, which often creates numerical difficulties on realistic datasets. We hence use Theorem \ref{cor2.3}'s result (\ref{bigtail}) instead, which gives an accurate approximation for sufficiently large $\Lambda/\sqrt{2}e^{\widehat{\bar{H}}_{t}}$. We therefore use the estimator
\begin{equation}
\widehat{\Delta}\left(\Lambda\right)=\text{argmin}_{\Delta}\left \{\left|\sum_{t=1}^{T}\left(\mathds{1}\left(\left|\varepsilon_{t}\right| \geq \Lambda\right)-\frac{\sqrt{2}^{1/\Delta}}{2\sqrt{\pi}} \Gamma \left(\frac{1+1/\Delta}{2}\right) \exp\left(\frac{\widehat{\bar{H}}_{t}}{\Delta}\right)\Lambda^{-1/\Delta} \right)\right |\right\}
\label{delest}
\end{equation}
In the simulation and empirical sections below, we search for (\ref{delest})'s $\widehat{\Delta}$ over the range $[.01,1]$ with a precision of .01.

\section{Simulation: $\Delta$ Estimation}
\label{sec: delta estimation}
In this section we present results of the estimation procedure described in the previous section for simulations of the process (\ref{process})-(\ref{laplace}). We run 1000 simulations each, with sample sizes of 625 and 1250, for $\Delta$ ranging from .05 to .50. The model used for (\ref{regmodel})'s $\widehat{\bar{H}}_{t}$ is a linear autoregressive (AR) model that uses the 10 previous values of (\ref{hest})'s $\widehat{H}_{t}$. This AR(10) model is calibrated using the Yule-Walker method. 

We present simulation results for two different, simple processes for $H_{t}$. The first is given by the AR(2) process 
\begin{equation}
H_{t}=.5H_{t-1}+.4H_{t-2}+h_{t} \label{simprocess1}
\end{equation} 
and the second is given by the AR(5) process
\begin{equation}
H_{t}=.05H_{t-1}+.05H_{t-2}+.25H_{t-3}+.2H_{t-4}+.35H_{t-5}+h_{t}. \label{simprocess2}
\end{equation}
The results are presented in Table \ref{ar}, where we give the averages and standard deviations of $\widehat{\Delta}$ over the 1000 simulations. For $\Lambda$, we choose $2\widehat{\sigma}_{\varepsilon}$, $3\widehat{\sigma}_{\varepsilon}$, and $4\widehat{\sigma}_{\varepsilon}$, where $\widehat{\sigma}_{\varepsilon}^{2}$ is $\varepsilon_{t}$'s sample variance. This gives a set of data-driven $\Lambda$ values for which (\ref{bigtail}) is an increasingly accurate approximation of (\ref{probconstraint})'s integral. 

\begin{table*}
\centering
\caption{Estimation results using (\ref{delest})'s $\widehat{\Delta}\left(\Lambda\right)$, averaged over 1000 simulations of $T$ samples of (\ref{process})-(\ref{laplace}) with $H_{t}$ given by (\ref{simprocess1}) and (\ref{simprocess2}). Model for (\ref{regmodel})'s $\widehat{\bar{H}}_{t}$ is AR(10).}
\label{ar}
\begin{tabular}{crrrrrrrrrrrrc}
\hline
$H_{t}=$(\ref{simprocess1})\\
\hline
$T=625-10$\\
\hline
$\Delta$ & $.05$ \hspace{1.5cm}& $.10$ \hspace{1.5cm}& $.15$ \hspace{1.5cm}& $.20$ \hspace{1.5cm}& $.25$ \hspace{1.5cm}& $.30$ \hspace{1.5cm}& $.35$ \hspace{1.5cm}& $.40$ \hspace{1.5cm}& $.45$ \hspace{1.5cm}& $.50$ 
\\
\hline
avg. $\widehat{\Delta}\left(2\widehat{\sigma}_{\varepsilon}\right)$        & .28 & .27 & .28 & .31 & .35 & .39 & .43 & .47 & .50 & .55\\ 
avg. $\widehat{\Delta}\left(3\widehat{\sigma}_{\varepsilon}\right)$        & .14 & .16 & .20 & .26 & .32 & .37 & .41 & .47 & .51 & .56\\ 
avg. $\widehat{\Delta}\left(4\widehat{\sigma}_{\varepsilon}\right)$        & .10 & .13 & .19 & .26 & .31 & .36 & .42 & .47 & .51 & .56\\ 
std. dev. $\widehat{\Delta}\left(2\widehat{\sigma}_{\varepsilon}\right)$  & .02 & .03 & .04 & .05 & .06 & .07 & .08 & .09 & .11 & .13\\
std. dev. $\widehat{\Delta}\left(3\widehat{\sigma}_{\varepsilon}\right)$  & .02 & .03 & .06 & .07 & .09 & .10 & .11 & .13 & .14 & .15\\
std. dev. $\widehat{\Delta}\left(4\widehat{\sigma}_{\varepsilon}\right)$  & .03 & .05 & .07 & .08 & .09 & .11 & .12 & .13 & .14 & .16 \\ 
\hline
$T=1250-10$\\
\hline
$\Delta$ & $.05$ \hspace{1.5cm}& $.10$ \hspace{1.5cm}& $.15$ \hspace{1.5cm}& $.20$ \hspace{1.5cm}& $.25$ \hspace{1.5cm}& $.30$ \hspace{1.5cm}& $.35$ \hspace{1.5cm}& $.40$ \hspace{1.5cm}& $.45$ \hspace{1.5cm}& $.50$ 
\\
\hline
avg. $\widehat{\Delta}\left(2\widehat{\sigma}_{\varepsilon}\right)$        & .26 & .25 & .27 & .30 & .34 & .38 & .41 & .45 & .50 & .54\\ 
avg. $\widehat{\Delta}\left(3\widehat{\sigma}_{\varepsilon}\right)$        & .13 & .15 & .21 & .27 & .32 & .37 & .41 & .46 & .51 & .55\\ 
avg. $\widehat{\Delta}\left(4\widehat{\sigma}_{\varepsilon}\right)$        & .09 & .14 & .21 & .26 & .32 & .38 & .42 & .47 & .52 & .56\\ 
std. dev. $\widehat{\Delta}\left(2\widehat{\sigma}_{\varepsilon}\right)$  & .01 & .02 & .02 & .03 & .04 & .05 & .06 & .08 & .09 &.11\\
std. dev. $\widehat{\Delta}\left(3\widehat{\sigma}_{\varepsilon}\right)$  & .01 & .03 & .06 & .07 & .08 & .09 & .10 & .11 & .12 & .14\\
std. dev. $\widehat{\Delta}\left(4\widehat{\sigma}_{\varepsilon}\right)$  & .02 & .05 & .06 & .07 & .08 & .09 & .10 & .12 & .14 & .14 \\ 
\hline
$H_{t}=$(\ref{simprocess2})\\
\hline
$T=625-10$\\
\hline
$\Delta$ & $.05$ \hspace{1.5cm}& $.10$ \hspace{1.5cm}& $.15$ \hspace{1.5cm}& $.20$ \hspace{1.5cm}& $.25$ \hspace{1.5cm}& $.30$ \hspace{1.5cm}& $.35$ \hspace{1.5cm}& $.40$ \hspace{1.5cm}& $.45$ \hspace{1.5cm}& $.50$ 
\\
\hline
avg. $\widehat{\Delta}\left(2\widehat{\sigma}_{\varepsilon}\right)$        & .29 & .27 & .26 & .26 & .28 & .32 & .36 & .41 & .46 & .52\\ 
avg. $\widehat{\Delta}\left(3\widehat{\sigma}_{\varepsilon}\right)$        & .15 & .14 & .17 & .21 & .26 & .32 & .38 & .43 & .48 & .52\\ 
avg. $\widehat{\Delta}\left(4\widehat{\sigma}_{\varepsilon}\right)$        & .10 & .11 & .15 & .20 & .26 & .32 & .37 & .42 & .48 & .53\\ 
std. dev. $\widehat{\Delta}\left(2\widehat{\sigma}_{\varepsilon}\right)$  & .02 & .02 & .03 & .03 & .05 & .07 & .09 & .10 & .12 & .12\\
std. dev. $\widehat{\Delta}\left(3\widehat{\sigma}_{\varepsilon}\right)$  & .02 & .02 & .04 & .06 & .08 & .08 & .09 & .10 & .10 & .12\\
std. dev. $\widehat{\Delta}\left(4\widehat{\sigma}_{\varepsilon}\right)$  & .02 & .04 & .06 & .07 & .08 & .08 & .08 & .09 & .10 & .11\\ 
\hline
$T=1250-10$\\
\hline
$\Delta$ & $.05$ \hspace{1.5cm}& $.10$ \hspace{1.5cm}& $.15$ \hspace{1.5cm}& $.20$ \hspace{1.5cm}& $.25$ \hspace{1.5cm}& $.30$ \hspace{1.5cm}& $.35$ \hspace{1.5cm}& $.40$ \hspace{1.5cm}& $.45$ \hspace{1.5cm}& $.50$ 
\\
\hline
avg. $\widehat{\Delta}\left(2\widehat{\sigma}_{\varepsilon}\right)$        & .27 & .26 & .25 & .25 & .27 & .32 & .36 & .41 & .46 & .52\\ 
avg. $\widehat{\Delta}\left(3\widehat{\sigma}_{\varepsilon}\right)$        & .13 & .13 & .17 & .23 & .28 & .33 & .38 & .43 & .48 & .53\\ 
avg. $\widehat{\Delta}\left(4\widehat{\sigma}_{\varepsilon}\right)$        & .09 & .11 & .16 & .22 & .28 & .33 & .38 & .43 & .48 & .53\\ 
std. dev. $\widehat{\Delta}\left(2\widehat{\sigma}_{\varepsilon}\right)$  & .01 & .02 & .02 & .03 & .05 & .06 & .08 & .09 & .10 & .11\\
std. dev. $\widehat{\Delta}\left(3\widehat{\sigma}_{\varepsilon}\right)$  & .01 & .02 & .05 & .06 & .07 & .07 & .08 & .08 & .09 & .10\\
std. dev. $\widehat{\Delta}\left(4\widehat{\sigma}_{\varepsilon}\right)$  & .02 & .04 & .05 & .06 & .06 & .07 & .07 & .08 & .08 & .10 \\ 
\hline
\end{tabular}
\end{table*}

One can see in Table \ref{ar} that for $H_{t}$ given by (\ref{simprocess1}) (resp. (\ref{simprocess2})) that for $\Delta \leq .30$ (resp. .25), $\widehat{\Delta}\left(k \widehat{\sigma}_{\varepsilon}\right)$'s bias appears to decrease as $k$ increases, while for $\Delta>.30$ (resp. .25) the bias seems to remain constant or slightly increase with $k$. The estimator's variance appears to nearly always increase with $k$, which is likely due to the higher variance in probability estimates for larger deviations. It is also clear that the variance increases with $\Delta$, which could be related to the increased variance in $\widehat{\sigma}_{\varepsilon}$ for larger $\Delta$. Doubling the sample size decreased the estimators' standard deviations by .01 to .02, and had a similar effect on the some of the biases. Although the bias actually increased in many cases by increasing the sample size.

The estimator $\widehat{\Delta}\left(2 \widehat{\sigma}_{\varepsilon}\right)$'s bias is very large for small $\Delta$ and decreases greatly as $\Delta \rightarrow 1/2$. To understand this behavior, recall that (\ref{datserieswrittenout})'s correction terms are inversely proportional to $\Delta^2$, which makes Theorem \ref{cor2.3}'s approximation less accurate for smaller $\Delta$. Intuitively, this is because as $\Delta$ increases, (\ref{bigprob}) becomes more sharply peaked around the origin. Hence a given $\Lambda$, such as $2 \widehat{\sigma}_{\varepsilon}$, will become sufficiently far in the tails for (\ref{bigtail}) to be a good approximation to (\ref{probconstraint})'s integral, making (\ref{delest}) a more appropriate estimator.

Meanwhile for $k=3$ and $4$, $\widehat{\Delta}\left(k \widehat{\sigma}_{\varepsilon}\right)$'s bias appears to be much smaller than $\widehat{\Delta}\left(2 \widehat{\sigma}_{\varepsilon}\right)$'s, and minimal around $.10 \leq \Delta \leq .2$. The inherent upward bias in the $\widehat{\Delta}$'s is likely due to model error in $\widehat{\bar{H}}_{t}$, since decreased predictability of $H_{t}$ manifests as a higher empirical $\Delta$. This explains the increases in bias seen when doubling the sample size, because the AR(10) models for $\widehat{\bar{H}}_{t}$ overfitted the data less on the larger samples. 

The processes (\ref{simprocess1})-(\ref{simprocess2}) were selected for their simplicity and the superficial similarity between their short-term autocorrelation structure and those seen in financial data sets. In particular, (\ref{simprocess1}) and (\ref{simprocess2})'s corresponding $\left|\varepsilon_{t}\right|$'s appear to be slightly more strongly locally correlated, but otherwise their short-term autocorrelations resemble those seen in recent daily S\&P 500 and U.S. Dollar/Euro absolute log-returns. Hence these simulation results show that with a realistically sized sample and similar conditions to the financial application below, $\widehat{\Delta}\left(k\widehat{\sigma}_{\varepsilon}\right)$ with $k=3$ or $4$ can provide a computationally inexpensive and satisfactory estimate of $\Delta$, given a predictive model for $H_{t}$. Now that we have shown the effectiveness of Section \ref{sec: delta estimation}'s $\Delta$ estimation procedure, in the next two sections we will show the predictive capabilities of our modeling method using simulated and financial time series data.

\section{Simulation: Lorenz-Driven Volatility}
\label{sec: lorenz}

In this section we will apply our modeling method to a simulated nonlinear time series where the volatility is driven by the Lorenz system. Note that this makes the volatility non-stochastic but instead driven by a deterministic, nonlinear system. This simulation study will show that even when the underlying process is not defined by (\ref{process})-(\ref{laplace}), the modeling method we have presented can still deliver highly predictive, dynamic estimates of volatility and extreme event probabilities.

The Lorenz system is a well-known set of ordinary differential equations, originally formulated as a simplified model for atmospheric convection by Lorenz (1963). The system of equations, given by 
\begin{equation}
\frac{dx}{dt}=\sigma\left(y-x\right), \hspace{.1cm} \frac{dy}{dt}=x\left(\rho-z\right)-y, \hspace{.1cm} \frac{dz}{dt}=xy-\beta z, \label{lorenz}
\end{equation}
describes a two-dimensional fluid layer uniformly heated from below and cooled from above. The variables $x$, $y$, and $z$ are proportional to the rate of convection, and the horizontal and vertical temperature variation respectively, while $\sigma$, $\rho$, and $\beta$ are physical parameters. The Lorenz system is highly nonlinear and famously exhibits chaotic behavior, which means the system's long-term evolution is highly sensitive to its initial conditions. Here we set the parameters $\sigma=10$, $\rho=28$, and $\beta=8/3$, which generate the system's most well-known and chaotic behavior. We use the R package "deSolve" given in Soetaert et al. (2010) to create 10,000 samples from the discretized Lorenz system with the given parameter values and initial conditions of $x = 0$, $y = 1$, and $z = 1$.

\begin{figure}
\includegraphics[width=\linewidth]{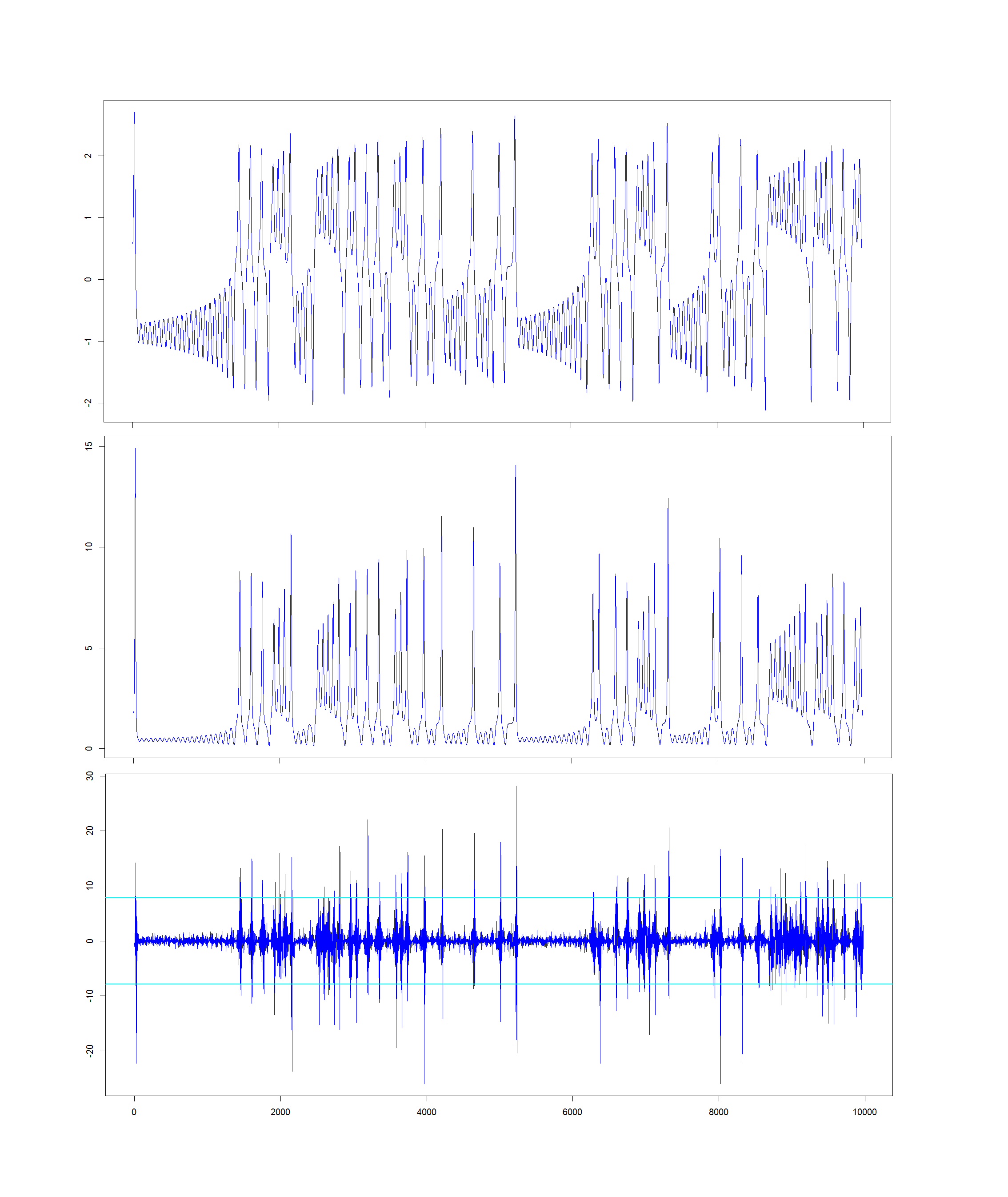}
\caption{Top Row: $H_{t}$ as defined in (\ref{lorenzh}) by the centered and de-scaled Lorenz output $x_{t}$, which is given by the discretized version of (\ref{lorenz}). Middle Row: The resulting volatility $\sigma_{t}=e^{H_{t}}$. Bottom Row: The resulting observable, heavy-tailed white noise $\varepsilon_{t}=\sigma_{t}z_{t}$ with $3\widehat{\sigma}_{\varepsilon}$ levels marked in Cyan.}
\label{fig: lorenz1}
\end{figure}

The Lorenz system enters into our simulated time series by driving the volatility. In particular, we set (\ref{volatility})'s $H_{t}$ as
\begin{equation}
H_{t}=\frac{x_{t}-\mu_{x}}{\sigma_{x}}, \label{lorenzh}
\end{equation}
with $x_{t}$ defined by the discretized version of (\ref{lorenz}). Thus, $H_{t}$ is given by the centered and de-scaled $x$ output from the Lorenz system. To create the observable, heavy-tailed white noise $\varepsilon_{t}$, we then multiply (\ref{lorenzh})'s $\exp\left(H_{t}\right)$ by $z_{t} \sim \mathcal{N}(0,1)$ \textit{i.i.d.} The Lorenz-driven $H_{t}$ with the resulting volatility $\sigma_{t}$ and time series $\varepsilon_{t}$ are shown in Fig. \ref{fig: lorenz1}. Note the clear bursts of high volatility and extreme fluctuations. We will use our methodology to predict $\varepsilon_{t}$'s dynamically changing volatility and extreme event probabilities.

To make predictions from the observable time series $\varepsilon_{t}$, we build a sparse, linear, AR model for (\ref{regmodel})'s $\widehat{\bar{H}}_{t}$ using the 20 previous values of (\ref{hest})'s $\widehat{H}_{t}$. Hence (\ref{regmodel})'s set of explanatory variables is simply $\left\{\widehat{H}_{t-1},...,\widehat{H}_{t-20}\right\}$. Because of high correlation and potential multicollinearity in this set of explanatory variables, we first apply Principal Component Analysis (PCA) to the training data to create the uncorrelated Principal Components (PCs) $\phi^{(1)}_{t},...,\phi^{(20)}_{t}$. We then use Tibshirani's (1996) LASSO regression and Friedman et al.'s (2010) cyclic coordinate descent to give the model 
\begin{equation}
\widehat{\bar{H}}_{t}=\widehat{\beta}_{0}+\sum_{n=1}^{20}\widehat{\beta}_{n}\phi^{(n)}_{t}, \label{entmodel}
\end{equation}
where 
\begin{equation}
\widehat{\vec{\beta}}=\text{argmin}_{\vec{\beta}}\left\{\frac{1}{2T}\sum_{t=11}^{T+10}\left|\widehat{H}_{t}-\beta_{0}-\sum_{n=1}^{20}\beta_{n}\phi^{(n)}_{t}\right|^{2}+\lambda\sum_{n=1}^{20}|\beta_{n}|\right\}, \label{beta}
\end{equation}
and $\lambda$ minimizes $\widehat{\bar{H}}_{t}$'s mean absolute 10-fold cross-validation error. This procedure, which can be called L1-penalized PC Regression, gives a sparse AR model for $\widehat{\bar{H}}_{t}$. We will use this simple model with (\ref{delest})'s $\widehat{\Delta}=\widehat{\Delta}\left(4\widehat{\sigma}_{\varepsilon}\right)$ to give an estimate of $\varepsilon_{t}$'s next time-step volatility $\bar{\sigma}_{\varepsilon,t}=\sqrt{E\left\{\varepsilon_{t}^{2}|\mathcal{F}_{t-1}\right\}}$ as well as a dynamic estimate of the probability that $\left|\varepsilon_{t}\right| \geq 3\widehat{\sigma}_{\varepsilon}$.

We use the moment result (\ref{processmoment}) to give the following model for the next time-step volatility:
\begin{equation}
\widehat{\bar{\sigma}}_{\varepsilon,t}=\sqrt{E\left\{\varepsilon_{t}^{2}|\widehat{\bar{H}}_{t}, \widehat{\Delta}\right\}}=\exp\left(\widehat{{\bar{H}}}_{t}\right)\sqrt{\frac{1}{1-4\widehat{\Delta}^{2}}}. \label{rvmodel}
\end{equation}
We also use the asymptotic approximation (\ref{bigtail})-(\ref{lamsub}) to give the dynamic probability estimate
\begin{equation}
\widehat{P}\left\{|\varepsilon_{t}| \geq 3 \widehat{\sigma}_{\varepsilon}|\widehat{\bar{H}}_{t}, \widehat{\Delta}\right\}=\frac{\sqrt{2}^{1/\widehat{\Delta}}}{2\sqrt{\pi}}\Gamma \left(\frac{1+1/\widehat{\Delta}}{2}\right) \exp\left(\frac{\widehat{\bar{H}}_{t}}{\widehat{\Delta}}\right)\left(3 \widehat{\sigma}_{\varepsilon}\right)^{-1/\widehat{\Delta}}. \label{probmodel}
\end{equation}
To measure the predictive capability of (\ref{rvmodel}) we can simply use the empirical correlation
\begin{equation}
\rho_{|\varepsilon|,\widehat{\sigma}}=\text{corr}\left\{\left|\varepsilon_{t}\right|, \widehat{\bar{\sigma}}_{\varepsilon,t}     \right\}. \label{volcorr}
\end{equation}

To measure the predictiveness of (\ref{probmodel}), we use it to create a classifier. We begin by noting that the stationary probability that $\left|\varepsilon_{t}\right| \geq 3\widehat{\sigma}_{\varepsilon}$ under a Gaussian distribution is $\approx .0027$. We then use (\ref{probmodel}) to define a binary classifier $\xi_{t}$ with the form 
\begin{equation}
\xi_{t}=\left\{\begin{matrix} 1;\hspace{.3cm} \widehat{P}\left\{|\varepsilon_{t}| \geq 3 \widehat{\sigma}_{\varepsilon}|\widehat{\bar{H}}_{t}, \widehat{\Delta}\right\} \geq 5\times .0027

\\ 
0;\hspace{.3cm} \widehat{P}\left\{|\varepsilon_{t}| \geq 3 \widehat{\sigma}_{\varepsilon}|\widehat{\bar{H}}_{t}, \widehat{\Delta}\right\} < 5\times .0027

\end{matrix}\right. \label{probclass}
\end{equation} 
Hence $\xi_{t}$ identifies times when the probability that $|\varepsilon_{t}| \geq 3 \widehat{\sigma}_{\varepsilon}$ exceeds 5 times the stationary Gaussian probability. We then calculate $\xi_{t}$'s sensitivity (Sn.) and specificity (Sp.), which are defined here as
\begin{equation}
\text{Sn.}=\frac{\left|\left\{\varepsilon_{t}\hspace{.2cm} \text{s.t.} \hspace{.2cm} |\varepsilon_{t}| \geq 3 \widehat{\sigma}_{\varepsilon} \wedge \xi_{t}=1\right\}\right|}{\left|\left\{\varepsilon_{t} \hspace{.2cm} \text{s.t.} \hspace{.2cm} |\varepsilon_{t}| \geq 3 \widehat{\sigma}_{\varepsilon} \right\} \right|} \label{sens}
\end{equation}
and
\begin{equation}
\text{Sp.}=\frac{\left|\left\{\varepsilon_{t}\hspace{.2cm} \text{s.t.} \hspace{.2cm} |\varepsilon_{t}| < 3 \widehat{\sigma}_{\varepsilon} \wedge \xi_{t}=0\right\}\right|}{\left|\left\{\varepsilon_{t} \hspace{.2cm} \text{s.t.} \hspace{.2cm} |\varepsilon_{t}| < 3 \widehat{\sigma}_{\varepsilon} \right\} \right|}. \label{spec}
\end{equation}
The measure (\ref{sens}) gives the percentage of 3 standard deviation events predicted by $\xi_{t}$, while (\ref{spec}) gives the percentage of non-3 standard deviation events predicted by $\xi_{t}$. We evaluate our modeling by doing three different sets of training and testing exercises for several different ratios of training sample size to testing sample size. We run 100 tests each and list averages of (\ref{delest})'s $\widehat{\Delta}\left(4\widehat{\sigma}_{\varepsilon}\right)$, (\ref{volcorr})'s $\rho_{|\varepsilon|,\widehat{\sigma}}$, and (\ref{sens})-(\ref{spec}) in Table \ref{traintestlorenz}. 

\begin{table*}
\centering
\caption{Lorenz-Driven Volatility Testing Results: Averages ($\pm$Standard Deviation)}
\label{traintestlorenz}
\begin{tabular}{crrrrrrrrrrrrc}
\hline
$N_{\text{Train}}/N_{\text{Test}}$ \hspace{.1cm}& avg. $\widehat{\Delta}\left(4 \widehat{\sigma}_{\varepsilon}\right)$ \hspace{.1cm}&  avg. $\rho_{|\varepsilon|,\widehat{\sigma}}$  \hspace{.1cm}& avg. Sn.  \hspace{.1cm}& avg. Sp.
\\
\hline
2,994/6,987 (30/70) \hspace{.1cm}& .359 ($\pm .008$) \hspace{.1cm}& .604 ($\pm .001$) \hspace{.1cm}&  .955 ($\pm .002$) \hspace{.1cm}& .745 ($\pm .008$)\\ 
3,992/5,989 (40/60) \hspace{.1cm}& .345 ($\pm .006$) \hspace{.1cm}& .614 ($\pm .000$) \hspace{.1cm}&  .934 ($\pm .003$) \hspace{.1cm}& .781 ($\pm .005$)\\ 
4,990/4,991 (50/50) \hspace{.1cm}& .347 ($\pm .005$) \hspace{.1cm}& .608 ($\pm .000$) \hspace{.1cm}&  .958 ($\pm .003$) \hspace{.1cm}& .736 ($\pm .006$)\\ 
\hline
\end{tabular}
\end{table*}

Note from Table \ref{traintestlorenz} that the modeling is highly predictive of volatility and extreme events in $\varepsilon_{t}$. In particular, (\ref{probmodel}) and (\ref{probclass})'s naive classifier $\xi_{t}$ is capable of predicting 93\% to 96\% of 3 standard deviation events out-of-sample. The classifier does this while still maintaining a high specificity of 73\% to over 78\%. We also note that (\ref{rvmodel})'s $\widehat{\bar{\sigma}}_{\varepsilon,t}$ achieves over 60\% correlation with $\left|\varepsilon_{t} \right|$ out-of-sample, which shows the modeling's ability to predict fluctuations in volatility. This predictive performance is achieved with training data sets that are smaller than the testing data sets. 

It is important to note that this simulated time series is a considerable departure from the process defined in (\ref{process})-(\ref{laplace}), as well as from the linear autoregressive model for $H_{t}$ defined in (\ref{entmodel})-(\ref{beta}). The volatility in this time series is not stochastic, but is instead driven by a deterministic, nonlinear, and chaotic system that involves unobserved variables. Nevertheless, our modeling method can produce highly predictive results. This suggests that this methodology could be useful for modeling a wide variety of nonlinear and heavy-tailed time series. Since Litimi et al. (2019) have recently demonstrated that chaos is likely to be present in financial market volatility, these simulation results are potentially directly relevant to financial applications. In the next section we will present an empirical application to financial market data where the modeling method can achieve similar predictive performance to that shown in Table \ref{traintestlorenz}.

\section{Empirical Application}
\label{sec: empirical}
In this section we apply our modeling and estimation procedure to daily log-returns of the S\&P 500 Index (SPX). We first consider a small sample of recent data from the last 5 years before next considering a much larger sample of data from the last 29 years. For both data sets we build a sparse, linear regression model for (\ref{regmodel})'s $\widehat{\bar{H}}_{t}$ using two sets of covariates. The first set is the 10 previous values of (\ref{hest})'s $\widehat{H}_{t}$, $\left\{\widehat{H}_{t-1},...,\widehat{H}_{t-10}\right\}$. The second set is the 10 previous values of $\log\left(\text{VIX}_{t}\right)$, $\left\{\log\left(\text{VIX}_{t-1}\right),...,\log\left(\text{VIX}_{t-10}\right)\right\}$, where VIX refers to the Chicago Board Options Exchange (CBOE) Volatility Index, which is an implied volatility measure computed from SPX option prices. It is perhaps the most well-known and widely disseminated measure of implied volatility for the S\&P 500 Index. 

Similarly to the previous section, because of potential multicollinearity in the model's set of explanatory variables 
\begin{equation}
\left\{\widehat{H}_{t-1},...,\widehat{H}_{t-10},\log \left(\text{VIX}_{t-1}\right),...,\log \left(\text{VIX}_{t-10}\right)\right\}, \label{evars}
\end{equation}
we again apply PCA to the training data to give the PCs $\phi^{(1)}_{t},...,\phi^{(20)}_{t}$. We also again use LASSO regression, giving a model of the form (\ref{entmodel})-(\ref{beta}). We use the model for $\widehat{\bar{H}}_{t}$ to similarly give the dynamic volatility and probability estimates (\ref{rvmodel})-(\ref{probmodel}) as well as the classifier (\ref{probclass}).

\subsection{2014 to 2019}

We first consider approximately 5 years of data from February 27, 2014 to February 14, 2019, giving 1,250 total samples. We evaluate our modeling using backtesting. In particular we do two sets of backtests. In the first backtesting exercise, we train the model on the first 50\% of the data, or 625 samples through August 18, 2016. We then test the model's performance on the second 50\% or 625 samples of the data. In the second backtesting exercise we train the model on the first 60\% of the data, or 750 samples through February 17, 2017. We test the model's performance on the second 40\% or 500 samples of the data. We run both backtests 100 times each and list averages of (\ref{delest})'s $\widehat{\Delta}\left(4\widehat{\sigma}_{\varepsilon}\right)$, (\ref{volcorr})'s $\rho_{|\varepsilon|,\widehat{\sigma}}$, and (\ref{sens})-(\ref{spec}) in Table \ref{traintest}.

\begin{table*}
\centering
\caption{SPX Backtesting Results: Averages ($\pm$Standard Deviation)}
\label{traintest}
\begin{tabular}{crrrrrrrrrrrrc}
\hline
2014 to 2019\\
\hline
$N_{\text{Train}}/N_{\text{Test}}$ \hspace{.1cm}& avg. $\widehat{\Delta}\left(4 \widehat{\sigma}_{\varepsilon}\right)$ \hspace{.1cm}&  avg. $\rho_{|\varepsilon|,\widehat{\sigma}}$  \hspace{.1cm}& avg. Sn.  \hspace{.1cm}& avg. Sp.
\\
\hline
625/625 (50/50) \hspace{.1cm}& .33 ($\pm .03$) \hspace{.1cm}& .39 ($\pm .02$) \hspace{.1cm}&  .78 ($\pm .00$) \hspace{.1cm}& .74 ($\pm .04$)\\ 
750/500 (60/40) \hspace{.1cm}& .28 ($\pm .02$) \hspace{.1cm}& .47 ($\pm .01$) \hspace{.1cm}& .85 ($\pm .05$) \hspace{.1cm}& .76 ($\pm .02$)\\ 
\hline
1990 to 2019\\
\hline
$N_{\text{Train}}/N_{\text{Test}}$ \hspace{.1cm}& avg. $\widehat{\Delta}\left(4 \widehat{\sigma}_{\varepsilon}\right)$ \hspace{.1cm}&  avg. $\rho_{|\varepsilon|,\widehat{\sigma}}$  \hspace{.1cm}& avg. Sn.  \hspace{.1cm}& avg. Sp.
\\
\hline
3,666/3,666 (50/50) \hspace{.1cm}& .206 ($\pm .005$) \hspace{.1cm}& .575 ($\pm .000$) \hspace{.1cm}&  .893 ($\pm .010$) \hspace{.1cm}& .805 ($\pm .003$)\\ 
4,399/2,933 (60/40) \hspace{.1cm}& .199 ($\pm .003$) \hspace{.1cm}& .584 ($\pm .000$) \hspace{.1cm}& .913 ($\pm .000$) \hspace{.1cm}& .733 ($\pm .003$)\\ 
\hline
\end{tabular}
\end{table*}

\subsection{1990 to 2019}

We now consider approximately 29 years of data from January 17, 1990 to March 1, 2019, giving 7,332 total samples. We again perform two sets of backtests. In the first backtesting exercise, we train the model on the first 50\% of the data, or 3,666 samples through August 4, 2004. We then test the model's performance on the second 50\% or 3,666 samples of the data. In the second backtesting exercise we train the model on the first 60\% of the data, or 4,399 samples through July 3, 2007. We test the model's performance on the second 40\% or 2,933 samples of the data. We again run both backtests 100 times each and list averages of (\ref{delest})'s $\widehat{\Delta}\left(4\widehat{\sigma}_{\varepsilon}\right)$, (\ref{volcorr})'s $\rho_{|\varepsilon|,\widehat{\sigma}}$, and (\ref{sens})-(\ref{spec}) in Table \ref{traintest}. 

\subsection{Discussion}

It is clear from Table \ref{traintest} that the modeling is highly predictive of extreme events in daily fluctuations of the S\&P 500 Index. (\ref{probmodel}) and (\ref{probclass})'s naive classifier $\xi_{t}$ can predict 85\% of 3 standard deviation events out-of-sample, and was observed to predict as much as 91\% of such events. The classifier does this while maintaining a specificity of 76\% on average, with an observed range of 71\% to 80\%. We also note that (\ref{rvmodel})'s $\widehat{\bar{\sigma}}_{\varepsilon,t}$ achieves nearly 50\% correlation with $\left|\varepsilon_{t} \right|$ out-of-sample, showing the model's ability to predict fluctuations in market volatility. These results are achieved using very small amounts of training data.

On the larger data set, $\xi_{t}$ shows a still better balance of sensitivity and specificity, predicting 89\% to 91\% of 3 standard deviation events on average, while maintaining a specificity of 73\% to 81\%. Additionally, the correlation between $\widehat{\bar{\sigma}}_{\varepsilon,t}$ and $\left|\varepsilon_{t} \right|$ reaches 58\%, indicating even higher capability for predicting fluctuations in volatility. Note that the larger-sample testing data included the exceptionally high volatility periods of the financial crisis. 

We note that, by Theorem \ref{cor2.3}, the approximation (\ref{probmodel}) diverges beyond unity for high values of $\widehat{\bar{H}}_{t}$. This exceedance beyond unity occasionally occurs in these empirical applications as well as in the previous section's application to Lorenz-driven volatility. In such extreme environments, where $\widehat{\bar{H}}_{t}$ is very large, (\ref{probmodel}) cannot be relied upon to give accurate probability estimates. Rather, (\ref{bigtail})-(\ref{lamsub}) and corresponding estimates with the form of (\ref{probmodel}) can be used as early warning tools for upcoming extreme events.

\begin{figure}
\includegraphics[width=\linewidth]{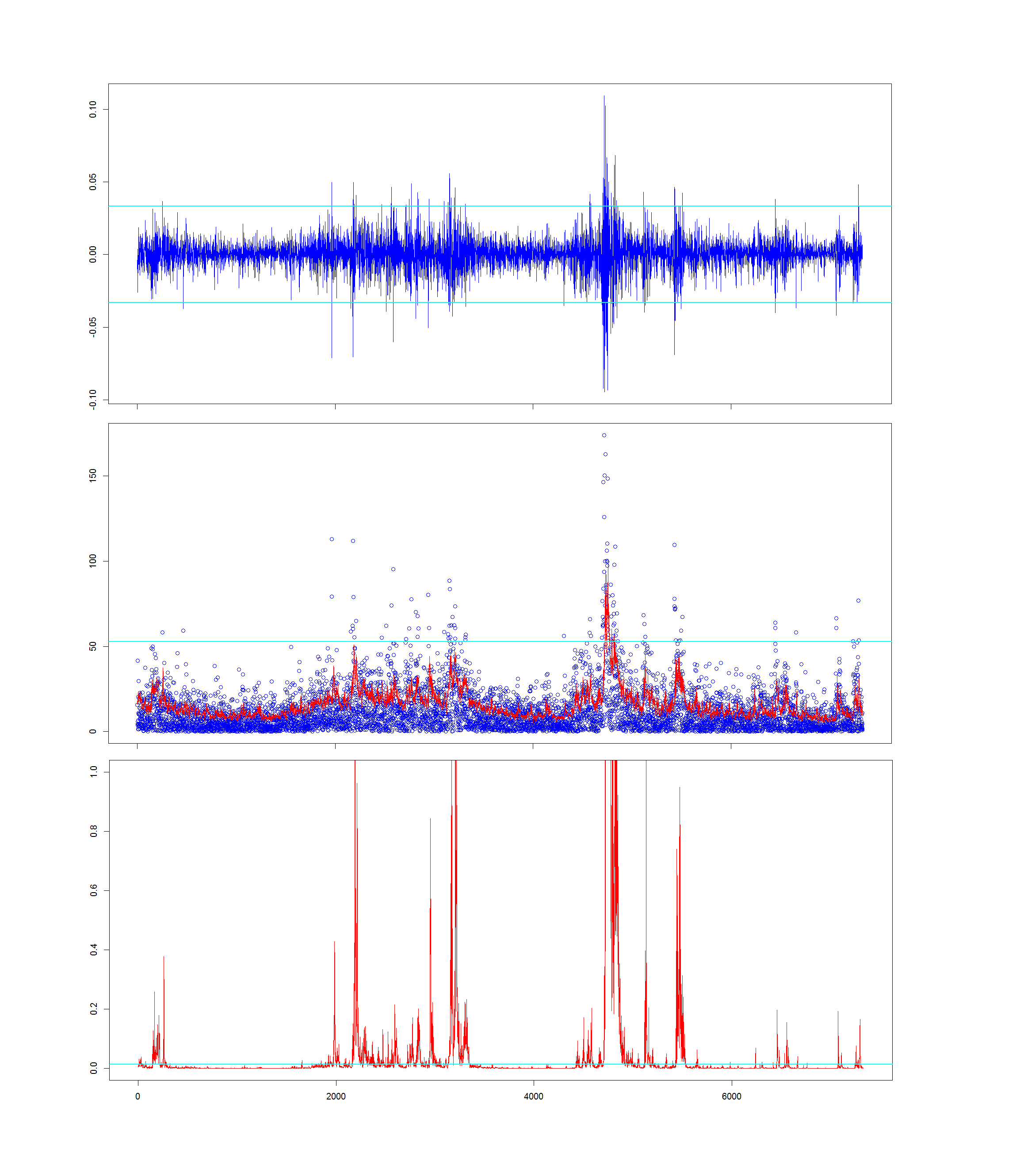}
\caption{Top Row: SPX log-returns $\varepsilon_{t}$ with $3 \widehat{\sigma}_{\varepsilon}$ levels marked in Cyan. Middle Row: Annualized $\left|\varepsilon_{t}\right|$ in Blue and (\ref{rvmodel})'s $\widehat{\bar{\sigma}}_{\varepsilon,t}$ in Red with $3\widehat{\sigma}_{\varepsilon}$ marked in Cyan. Bottom Row: (\ref{probmodel})'s $\widehat{P}\left\{|\varepsilon_{t}| \geq 3 \widehat{\sigma}_{\varepsilon}|\widehat{\bar{H}}_{t}, \widehat{\Delta}\right\}$ in Red with the threshold value for (\ref{probclass})'s $\xi_{t}$ marked in Cyan.  (\ref{delest})'s $\widehat{\Delta}\left(4\widehat{\sigma}_{\varepsilon}\right)$ gives the whole-sample result $\widehat{\Delta}_{\text{SPX}}=.20$.}
\label{fig: 3}
\end{figure}

We also note that since the model for $\widehat{\bar{H}}_{t}$ uses the 10 previous values of $\widehat{H}_{t}$ and $\log\left(\text{VIX}_{t}\right)$, it has a memory length of about two trading weeks. This relatively naive, short-memory modeling of $H_{t}$ could potentially be improved for this application, since many researchers have found empirical evidence of long-memory in SPX volatility (see Ding et al. 1993, Bollerslev and Mikkelsen 1996, Lobato and Savin 1998, Ray and Tsay 2000, Grau-Carles 2000). The modeling could also be adjusted to accomodate the asymmetric tails observed in stock market data, often called the "leverage effect" (see Black 1976, Christie 1982, Nelson 1991, Engle and Ng 1993). Overall, though, the study here shows that a simple, short-memory model for $H_{t}$ that uses only recent realized and implied volatility measures, with the parametrization (\ref{process})-(\ref{tailresult}) enables straightforward and predictive tail inference in financial time series. Plots of the SPX daily log-returns from 1990 to 2019 along with corresponding model outputs are given in Fig.\ref{fig: 3}.

\section{Conclusion}
We have presented a family of stochastic volatility models that enable dynamic tail inference in heavy-tailed time series. The family's conditional probabilistic structure allows for straightforward, effective, and computationally inexpensive estimation of model parameters and outcome probabilities. We have shown that this modeling formalism can be useful for predictive inference of dynamically changing extreme event probabilities in nonlinear and financial time series data. Current and future directions of research involve seeking formal results on model estimation, long-memory modeling for conditionally Laplace processes, accomodation of asymmetric volatility and tails, as well as time aggregation.

\section*{Acknowledgements}
The author thanks Prof. Gennady Samorodnitsky, Cornell University, and Prof. Richard Davis, Columbia University, for their comments on a very early version of this work in 2016, as well as Dr. Dobrislav Dobrev, Federal Reserve Board, for helpful comments on an early version of this manuscript at the 2017 10th Annual Society for Financial Econometrics (SoFiE) Conference. The author is also grateful to Dr. Altan Allawala, JPMorgan Chase \& Co., for his insightful comments. The author also thanks Prof. Richard Kleeman, Courant Institute of Mathematical Sciences, and Prof. Clifford Hurvich, NYU Stern, for many hours of motivating discussions, along with Prof. Fahad Saled, McGill University, and Franz Hinzen, NYU Stern, for encouragement and conversation on these topics. Finally, the author is very grateful to Prof. David Jablons, University of California San Francisco, for his generous encouragement and support. 

\appendix

\section{Appendix 1: Proofs}

\subsection{Proof of Lemma \ref{lem2.1}}
\begin{proof}
We first make the substitution $y=\log \left(y'/e^{\bar{H}_{t}} \right)$.
Then by (\ref{volatility}) and (\ref{laplace}), for $h_{t}\geq 0$,
$$P\left(\sigma_{t} \leq y'\right)=P\left(h_{t} \leq y \right)=\frac{1}{2}-\frac{1}{2}\exp \left(-\frac{\log \left(e^{-\bar{H}_{t}} y' \right)}{\Delta} \right)=\frac{1}{2}-\frac{1}{2}\left(\frac{e^{\bar{H}_{t}}}{ y'} \right)^{\frac{1}{\Delta}}$$
We then differentiate w.r.t. $y'$ and let $y'=\sigma_{t}$, which gives (\ref{loglaplace})'s result for $\sigma_{t} \geq \exp\left(\bar{H}_{t}\right)$. Next, by (\ref{volatility}), (\ref{laplace}), for $h_{t} <0$,
$$P\left(\sigma_{t} \leq y'\right)=P\left(h_{t} \leq y \right)=\frac{1}{2} \exp\left(\frac{\log \left(e^{-\bar{H}_{t}} y'\right)}{\Delta} \right)=\frac{1}{2} \left(\frac{y'}{e^{\bar{H}_{t}}}\right)^{\frac{1}{\Delta}}.$$
We again differentiate w.r.t. $y'$ and let $y'=\sigma_{t}$ to give (\ref{loglaplace})'s result for $0 \leq \sigma_{t} < \exp\left(\bar{H}_{t}\right)$.
\end{proof}

\subsection{Proof of Corollary \ref{cor2.1} }
\begin{proof}
We note from (\ref{loglaplace}) that for any $n\geq1$, the conditional expected value of $\sigma_{t}^{n}$ is equal to the sum of integrals 
$$
\hspace{-.1cm}E\left\{\sigma_{t}^{n}|\mathcal{F}_{t-1}\right\}=\frac{1}{2\Delta}\left(\exp\left(-\frac{\bar{H}_{t}}{\Delta}\right)\int_{0}^{\exp\left(\bar{H}_{t}\right)}\sigma_{t}^{1/\Delta+n-1}d\sigma_{t}+\exp\left(\frac{\bar{H}_{t}}{\Delta}\right)\int_{\exp\left(\bar{H}_{t}\right)}^{\infty}\sigma_{t}^{-1/\Delta+n-1}d\sigma_{t}\right),
$$
the second term of which diverges for $\Delta \geq 1/n$, while for $\Delta < 1/n$ we have after integration
\begin{equation}
E\left\{\sigma_{t}^{n}|\mathcal{F}_{t-1}\right\}=\frac{1}{2}\exp\left(n\bar{H}_{t}\right)\left(\frac{1}{1-n\Delta}+\frac{1}{1+n\Delta} \right)=\frac{\exp\left(n\bar{H}_{t}\right)}{1-n^{2}\Delta^{2}}, \label{volmoment}
\end{equation}
We next note that since $z_{t}$ is \textit{i.i.d.} and $z_{t}$ and $\sigma_{t}$ are independent, the conditional expectation $E\left\{\varepsilon_{t}^{n}|\mathcal{F}_{t-1}\right\}=Ez_{t}^{n}E\left \{\sigma_{t}^{n}|\mathcal{F}_{t-1}\right\}$. Since $z_{t}$ is standard normal, its $n$th moment is simply $(n-1)!!=(n-1)(n-3)(n-5)...1$. Therefore $E\left\{\varepsilon_{t}^{n}|\mathcal{F}_{t-1}\right\}=(n-1)!!E\left \{\sigma_{t}^{n}|\mathcal{F}_{t-1}\right\}$. Using (\ref{volmoment}) then gives (\ref{processmoment})'s result.
\end{proof}

\subsection{Proof of Theorem \ref{thm2.1}}
\begin{proof}
We note from Rohatgi (1976 p. 141) that since (\ref{process})'s $z_{t}$ and $\sigma_{t}$ are independent, the distribution of their product $\sigma_{t}z_{t}=\varepsilon_{t}$ is given by the formula 
\begin{equation}
p_{\varepsilon}(\varepsilon_{t}|\mathcal{F}_{t-1})=\int_{0}^{\infty} p_{\sigma}(\sigma_{t}|\mathcal{F}_{t-1})p_{z}\left(\frac{\varepsilon_{t}}{\sigma_{t}}\right)\frac{1}{|\sigma_{t}|}d\sigma_{t}. \label{productformula}
\end{equation}
We then substitute the standard normal density for $p_{z}$ and (\ref{loglaplace}) for $p_{\sigma}$ into (\ref{productformula}) to give 
\begin{eqnarray}
p_{\varepsilon}(\varepsilon_{t}|\mathcal{F}_{t-1})=\frac{1}{2 \Delta \sqrt{2 \pi}}\exp\left(-\frac{\bar{H}_{t}}{\Delta}\right)\int_{0}^{\exp\left(\bar{H}_{t}\right)}\sigma_{t}^{1/\Delta-2}e^{-\frac{\varepsilon_{t}^{2}}{2\sigma_{t}^{2}}}d\sigma_{t} \label{gottasimplify} \\ \nonumber +\frac{1}{2 \Delta \sqrt{2 \pi}}\exp\left(\frac{\bar{H}_{t}}{\Delta}\right)\int_{\exp\left(\bar{H}_{t}\right)}^{\infty}\sigma_{t}^{-1/\Delta-2}e^{-\frac{\varepsilon_{t}^{2}}{2\sigma_{t}^{2}}}d\sigma_{t}
\end{eqnarray}
We note from (\ref{upgamma}) that the first integral in (\ref{gottasimplify}) can be written in the following form after simplifying
\begin{equation}
\frac{\left(\sqrt{2}e^{\bar{H}_{t}}\right)^{-1/\Delta}}{4\sqrt{\pi} \Delta} \left|\varepsilon_{t} \right|^{1/\Delta-1}\left[\Gamma \left(\frac{1-1/\Delta}{2},\frac{\varepsilon_{t}^{2}}{2 \sigma_{t}^{2}}\right)_{\sigma_{t}=e^{\bar{H}_{t}}}-\Gamma \left(\frac{1-1/\Delta}{2},\frac{\varepsilon_{t}^{2}}{2 \sigma_{t}^{2}}\right)_{\sigma_{t}=0} \right]\label{i1}
\end{equation}
To simplify (\ref{i1}) further, we note from Abramowitz and Stegun (1965 p. 263) that the upper incomplete gamma function satisfies
\begin{equation}
\Gamma(a,b) \sim b^{a-1}e^{-b} \label{upgammaasympt}
\end{equation}
as $b \rightarrow \infty$. Letting $(1-1/\Delta)/2=a$ and $\varepsilon_{t}^{2}/2 \sigma_{t}^{2}=b$ in (\ref{upgammaasympt}) shows that the second term in (\ref{i1}) vanishes. Final simplification of (\ref{i1}) gives the result
\begin{equation}
\frac{1}{4 \sqrt{\pi} \Delta} \left(\sqrt{2}e^{\bar{H}_{t}}\right)^{-1/\Delta}\Gamma \left(\frac{1-1/\Delta}{2},\frac{\varepsilon_{t}^{2}}{2e^{2\bar{H}_{t}}}\right)\left|\varepsilon_{t} \right|^{1/\Delta-1}. \label{term1}
\end{equation}
The definition (\ref{upgamma}) can be used again to write the second integral in (\ref{gottasimplify}) as
\begin{equation}
\frac{\left(\sqrt{2}e^{\bar{H}_{t}}\right)^{1/\Delta}}{4 \sqrt{\pi} \Delta}\left|\varepsilon_{t} \right|^{-1/\Delta-1}\left[\Gamma\left(\frac{1+1/\Delta}{2}, \frac{\varepsilon_{t}^{2}}{2 \sigma_{t}^{2}} \right)_{\sigma_{t}=\infty} -\Gamma\left(\frac{1+1/\Delta}{2}, \frac{\varepsilon_{t}^{2}}{2 \sigma_{t}^{2}} \right)_{\sigma_{t}=e^{\bar{H}_{t}}}\right]
\label{i2}
\end{equation}
after simplifying. The first upper incomplete gamma function term in (\ref{i2}) is simply $\Gamma\left((1+1/\Delta)/2\right)$. We then note that 
\begin{equation}
\Gamma(a)-\Gamma\left(a,b\right)=\gamma\left(a,b\right).
\label{gammaidentity}
\end{equation}
Using this identity with (\ref{i2}) and simplifying gives
\begin{equation}
\frac{1}{4 \sqrt{\pi} \Delta}\left(\sqrt{2}e^{\bar{H}_{t}}\right)^{1/\Delta} \gamma \left(\frac{1+1/\Delta}{2},\frac{\varepsilon_{t}^{2}}{2e^{2\bar{H}_{t}}} \right)\left|\varepsilon_{t} \right|^{-1/\Delta-1}. \label{term2}
\end{equation}
Adding (\ref{term1}) and (\ref{term2}) gives the result in (\ref{bigprob}).
\end{proof}

\subsection{Proof of Corollary \ref{cor2.2}}
\begin{proof}
We first note the identity shown in Jameson (2016, 2017)
\begin{equation}
\Gamma\left(a,b\right)\sim -\frac{b^{a}}{a} \label{upgamma0}
\end{equation}
as $b\rightarrow 0$ for $a<0$. For $\Delta<1$, $(1-1/\Delta)/2<0$, and hence using (\ref{upgamma0}) and rewriting gives
\begin{equation}
\Gamma\left(\frac{1-1/\Delta}{2}, \frac{\varepsilon_{t}^{2}}{2e^{2\bar{H}_{t}}}\right)\sim \frac{2\Delta}{1-\Delta}\left(\frac{\left|\varepsilon_{t}\right|}{\sqrt{2}e^{\bar{H}_{t}}}\right)^{1-1/\Delta}
\label{near0res1}
\end{equation}
as $\left|\varepsilon_{t}\right| \rightarrow 0$. We next note the identity 
\begin{equation}
\gamma\left(a,b\right)\sim \frac{b^{a}}{a} \label{lowgamma0}
\end{equation}
as $b\rightarrow 0$. Using (\ref{lowgamma0}) and rewriting then gives 
\begin{equation}
\gamma \left(\frac{1+1/\Delta}{2}, \frac{\varepsilon_{t}^{2}}{2e^{2\bar{H}_{t}}}\right) \sim \frac{2\Delta}{1+\Delta}\left(\frac{\left|\varepsilon_{t}\right|}{\sqrt{2}e^{\bar{H}_{t}}}\right)^{1+1/\Delta}
\label{near0res2}
\end{equation}
as $\left|\varepsilon_{t}\right|\rightarrow 0$. Substituting (\ref{near0res1}) and (\ref{near0res2}) into (\ref{bigprob}) and cancelling terms gives 
$$
p_{\varepsilon}\left(\varepsilon_{t}|\mathcal{F}_{t-1}\right) \sim \frac{1}{4}\sqrt{\frac{2}{\pi}}\frac{1}{e^{\bar{H}_{t}}}\left(\frac{1}{1-\Delta}+\frac{1}{1+\Delta}\right)
$$ 
as $\left|\varepsilon_{t}\right|\rightarrow 0$, which can be simplified to give the result (\ref{near0res}).
\end{proof}

\subsection{Proof of Theorem \ref{closedformthm}}
\begin{proof}
We first make the substitution
\begin{equation}
b=\frac{\varepsilon_{t}}{\sqrt{2}e^{\bar{H}_{t}}}
\end{equation}
in (\ref{bigprob})'s integral from $\Lambda$ to $\infty$. Multiplying by 2, cancelling terms, and using the notation (\ref{lamsub}) then gives 
\begin{eqnarray}
P\left\{\left|\varepsilon_{t}\right|\geq \Lambda|\mathcal{F}_{t-1}\right\}=\frac{1}{2\sqrt{\pi}\Delta}\int_{\widetilde{\Lambda}_{t}}^{\infty}\Gamma\left(\frac{1-1/\Delta}{2},b^{2}\right)b^{1/\Delta-1}db \label{starting} \\ \nonumber +\frac{1}{2\sqrt{\pi}\Delta}\int_{\widetilde{\Lambda}_{t}}^{\infty}\gamma\left(\frac{1+1/\Delta}{2},b^{2}\right)b^{-1/\Delta-1}db. 
\end{eqnarray}
We begin with the first term of (\ref{starting}). We apply integration by parts with 
$$u=\Gamma\left(\frac{1-1/\Delta}{2},b^{2}\right), \hspace{.3cm} v=b^{1/\Delta}.$$
Then using the definition (\ref{upgamma}) for differentiation of $u$ and cancelling terms gives 
\begin{equation}\frac{1}{2\sqrt{\pi}}\left(\left[\Gamma\left(\frac{1-1/\Delta}{2},b^{2}\right)b^{1/\Delta}\right]_{b=\infty}-\left[\Gamma\left(\frac{1-1/\Delta}{2},b^{2}\right)b^{1/\Delta}\right]_{b=\widetilde{\Lambda}_{t}}+2\int_{\widetilde{\Lambda}_{t}}^{\infty}e^{-b^{2}}db\right).
\label{intterm1}
\end{equation}
Recall from (\ref{upgammaasympt}) that (\ref{intterm1})'s first term vanishes. We then proceed to (\ref{starting})'s second term, again applying integration by parts with 
$$u=\gamma\left(\frac{1+1/\Delta}{2},b^{2}\right), \hspace{.3cm} v=-b^{-1/\Delta}.$$
Then using (\ref{lowgamma}) for differentiation of $u$ and cancelling terms gives 
\begin{equation}\frac{1}{2\sqrt{\pi}}\left(-\left[\gamma\left(\frac{1+1/\Delta}{2},b^{2}\right)b^{-1/\Delta}\right]_{b=\infty}+\left[\gamma\left(\frac{1+1/\Delta}{2},b^{2}\right)b^{-1/\Delta}\right]_{b=\widetilde{\Lambda}_{t}}+2\int_{\widetilde{\Lambda}_{t}}^{\infty}e^{-b^{2}}db\right).
\label{intterm2}
\end{equation}
Note that (\ref{intterm2})'s first term also vanishes. Combining (\ref{intterm1}) and (\ref{intterm2}) and noting the definition (\ref{comperf}) gives the result (\ref{datclosedform}).
\end{proof}

\subsection{Proof of Theorem \ref{cor2.3}}
\begin{figure}
\includegraphics[width=\linewidth]{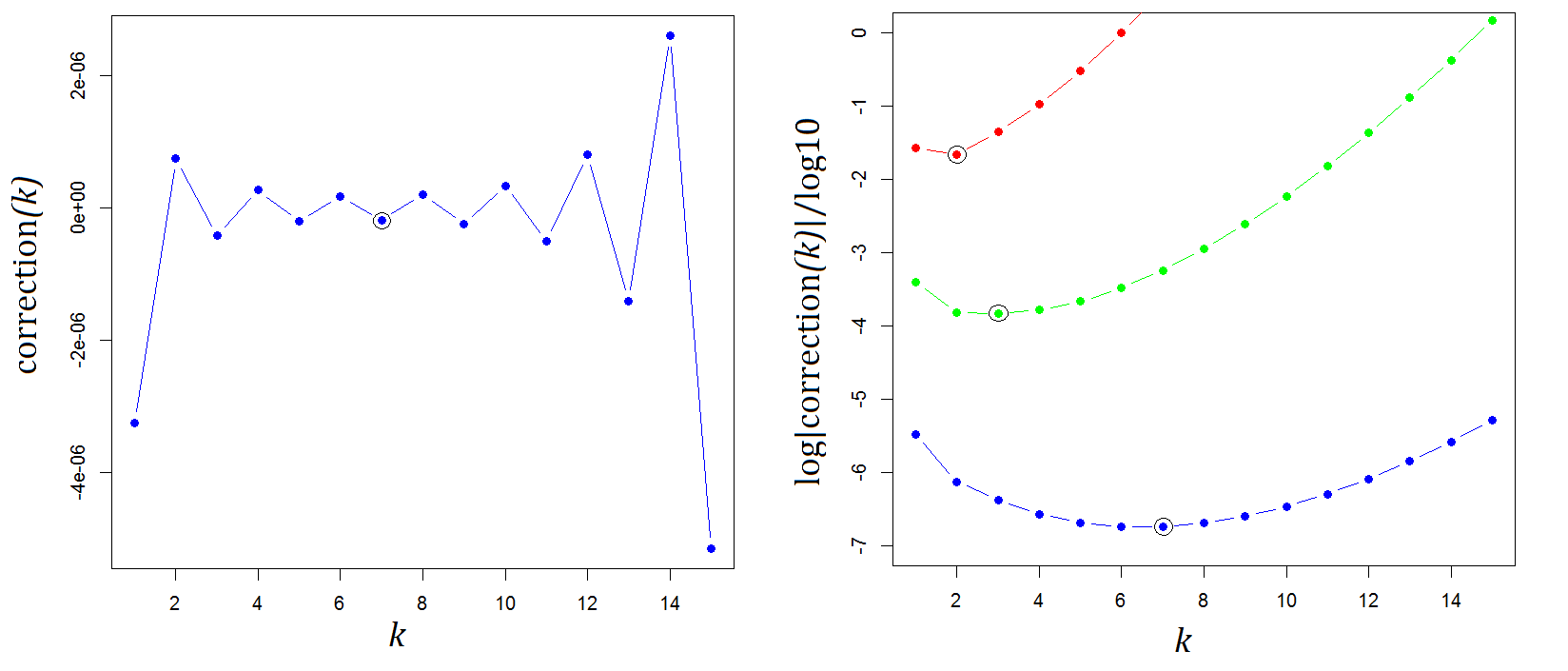}
\caption{Visualizations of (\ref{datseries})'s correction terms to (\ref{bigtail})-(\ref{lamsub}) for $k=1,...,15$ with $\Delta=1/4$. Left: A plot of the series' corrections for $\widetilde{\Lambda}_{t}=4/\sqrt{2}$. Right: Plots of the base-10 logarithm of the (absolute-valued) corrections for $\widetilde{\Lambda}_{t}=2/\sqrt{2}$ (Red), $\widetilde{\Lambda}_{t}=3/\sqrt{2}$ (Green), and $\widetilde{\Lambda}_{t}=4/\sqrt{2}$ (Blue). In both images the series' minimal correction terms before divergence begins are circled.}
\label{fig: series}
\end{figure}

\begin{proof}
We will use the following asymptotic expansion of (\ref{upgamma}), which can be derived from repeated integration by parts (see Digital Library of Mathematical Functions 8.11).
\begin{equation}
\Gamma \left(a,b\right)=b^{a-1}e^{-b}\left(1+\sum_{k=1}^{n-1}\frac{u_{k}(a)}{b^{k}}+O\left(b^{-n}\right)\right) \label{upgammaseries}
\end{equation}
as $b \rightarrow \infty$, where
\begin{equation}
u_{k}(a)=(-1)^{k}(1-a)_{k}=(a-k)(a-k+1)...(a-2)(a-1),\label{ukas}
\end{equation}
and $(.)_{k}$ denotes the rising factorial. We then make the following substitutions in (\ref{bigprob}),
\begin{equation}
a_{1}=\frac{1-1/\Delta}{2}, \hspace{.5cm} a_{2}=\frac{1+1/\Delta}{2}, \hspace{.5cm} b_{t}=\frac{\varepsilon_{t}^{2}}{2e^{2\bar{H}_{t}}}. \label{triplesubs}
\end{equation}
We then recall (\ref{gammaidentity}) and substitute (\ref{upgammaseries}) into (\ref{bigprob})'s two terms (\ref{term1}) and (\ref{term2}). Simplifying then gives the series representation of (\ref{bigprob}) as $b_{t}\rightarrow \infty$,
\begin{equation}
p_{\varepsilon}\left(\varepsilon_{t}|\mathcal{F}_{t-1}\right)=\frac{1}{4\Delta\sqrt{2\pi}e^{\bar{H}_{t}}}\left(b_{t}^{-a_{2}}\Gamma\left(a_{2}\right)+b_{t}^{-1}e^{-b_{t}}\sum_{k=1}^{n-1}\frac{u_{k}\left(a_{1}\right)-u_{k}\left(a_{2}\right)}{b_{t}^{k}}+O\left(b_{t}^{-n-1}e^{-b_{t}}\right)\right).\label{pdf}
\end{equation}
We next note that, after re-substituting for $b_{t}$ and $a_{2}$, the first term in (\ref{pdf}) is equal to
\begin{equation}
\frac{1}{4 \sqrt{\pi} \Delta}\left(\sqrt{2}e^{\bar{H}_{t}}\right)^{1/\Delta}\Gamma \left(\frac{1+1/\Delta}{2}\right) \left|\varepsilon_{t} \right|^{-1/\Delta-1}, \label{bigasympt}
\end{equation}
which is the limit of the term (\ref{term2}) as $\left|\varepsilon_{t}\right|\rightarrow \infty$. Multiplying (\ref{bigasympt}) by 2 and integrating from $\Lambda$ to $\infty$ gives the first term of (\ref{bigtail}) with (\ref{lamsub}). We then proceed to (\ref{pdf})'s second term, re-substituting for $b_{t}$ and cancelling factors to give
\begin{equation}
\frac{e^{\bar{H}_{t}}}{2\Delta\sqrt{2\pi}}\frac{e^{-\varepsilon_{t}^{2}/2e^{2\bar{H}_{t}}}}{\varepsilon_{t}^{2}}\sum_{k=1}^{n-1}\frac{u_{k}\left(a_{1}\right)-u_{k}\left(a_{2}\right)}{\varepsilon_{t}^{2k}}\left(2e^{2\bar{H}_{t}}\right)^{k}. \label{t2}
\end{equation}
We next note the following integral,
\begin{equation}
\int_{\Lambda}^{\infty} e^{-\varepsilon^{2}/2e^{2\bar{H}_{t}}}\varepsilon^{-2(k+1)}d\varepsilon=\frac{1}{2}\left(2e^{2\bar{H}_{t}}\right)^{-k-1/2}\Gamma\left(-k-\frac{1}{2},\frac{\Lambda^{2}}{2e^{2\bar{H}_{t}}}\right). \label{impint}
\end{equation}
So integrating (\ref{t2}) from $\Lambda$ to $\infty$, applying (\ref{impint}), and cancelling factors gives
\begin{equation}
\frac{1}{8\sqrt{\pi}\Delta}\sum_{k=1}^{n-1}\left[u_{k}\left(a_{1}\right)-u_{k}\left(a_{2}\right)\right]\Gamma\left(-k-\frac{1}{2},\frac{\Lambda^{2}}{2e^{2\bar{H}_{t}}}\right). \label{t2int}
\end{equation}
Because (\ref{pdf})'s remainder at any $n$ does not change sign over all $b_{t}$, we can similarly apply (\ref{impint}) to give the remainder term's integral. Finally, multiplying (\ref{pdf}) by 2 and integrating from $\Lambda$ to $\infty$, applying the integral of (\ref{bigasympt}), the result (\ref{t2int}), using the clearer notation for $u_{k}(a)$, and fully re-substituting gives the asymptotic expansion 
\begin{eqnarray}
\hspace{-1cm}P\left\{\left|\varepsilon_{t}\right| \geq \Lambda|\mathcal{F}_{t-1}\right\}=\frac{\sqrt{2}^{1/\Delta}}{2\sqrt{\pi}} \Gamma \left(\frac{1+1/\Delta}{2}\right) \exp\left(\frac{\bar{H}_{t}}{\Delta}\right)\Lambda^{-1/\Delta} \label{datseries} \\ \nonumber +\frac{1}{4\sqrt{\pi}\Delta}\sum_{k=1}^{n-1}\left(-1\right)^{k}\left[\left(\frac{1+1/\Delta}{2}\right)_{k}-\left(\frac{1-1/\Delta}{2}\right)_{k}\right]\Gamma\left(-k-\frac{1}{2},\frac{\Lambda^{2}}{2e^{2\bar{H}_{t}}}\right)\\ \nonumber + O\left(\Gamma \left(-n-\frac{1}{2},\frac{\Lambda^{2}}{2e^{2{\bar{H}_{t}}}}\right)\right). 
\end{eqnarray}
Re-applying (\ref{upgammaseries}) to (\ref{datseries})'s remainder term with $n=1$ gives the remainder term in (\ref{bigtail}) with (\ref{lamsub}). 
\end{proof}
The series in (\ref{pdf}) and (\ref{datseries}) are generally not convergent as $n \rightarrow \infty$. However, they can be truncated at low order $n$ to closely approximate (\ref{bigprob}) and its integral. This is because for $n \ll \infty$ the series' terms rapidly become very small for values of $\Lambda$ that are relatively large compared to $e^{\bar{H}_{t}}$. Visualizations of (\ref{datseries})'s asymptotic series are given in Fig. \ref{fig: series}.

\subsection{Proof of Proposition \ref{prop3.1}}
\begin{proof}
By (\ref{process}) and (\ref{volatility}), $\left|\varepsilon_{t}\right|=\exp\left(H_{t}\right)\left|z_{t}\right|$. Taking the logarithm gives
\begin{equation}
\log \left|\varepsilon_{t}\right|=H_{t}+\log \left|z_{t}\right|. \label{logprocess}
\end{equation}
Since $z_{t}$ and $\varepsilon_{t}$ are independent, $E\left\{\log\left|z_{t}\right||\log \left|\varepsilon_{t}\right|\right\}=E\log\left|z_{t}\right|$. Then since $z_{t}$ is standard normal, $z_{t}^{2} \sim \chi^{2}(1)$. It can then be noted, e.g., from Breidt et al. (1998) that 
\begin{equation}
E\log \left|z_{t}\right|=\frac{1}{2}E\log z_{t}^{2}=\frac{1}{2}\left(\log 2+\psi\left(\frac{1}{2}\right)\right)=\frac{-\log 2 -\gamma}{2}\approx -.6352, \label{logabsnorm}
\end{equation}
where $\psi(x)$ is the \textit{digamma function}
$$\psi(x)=\frac{d}{dx}\log\left(\Gamma(x)\right)=\frac{\Gamma ' (x)}{\Gamma(x)}$$
and $\psi(1/2)=-2\log2-\gamma$ (see Abramowitz and Stegun 1965 p. 258-259). Rearranging (\ref{logprocess}), taking the expected value $E\left \{H_{t}|\log \left|\varepsilon_{t}\right|\right\}$, and substituting (\ref{logabsnorm}) gives the result (\ref{prehest}).
\end{proof}

\section{Appendix 2: The $z_{t}\sim Lap(0,1)$ Case}

In this section we briefly show that very similar results to those in Appendix 1 for (\ref{process})-(\ref{laplace}) with $z_{t}\sim \mathcal{N}(0,1)$ can be derived for the case where $z_{t}$ is instead distributed according to the standard Laplace density 
\begin{equation}
p_{z}\left(z_{t}\right)=\frac{1}{2}\exp\left(-\left|z_{t}\right|\right), \label{zlap} 
\end{equation}
which gives $Ez_{t}=0$ and $E\left|z_{t}\right|=1$. We first note that an equivalent argument to that in A.2 can be made. We use the fact that (\ref{zlap})'s $Ez_{t}^{n}=n!$ with the result (\ref{volmoment}) to give
\begin{equation}
E\left\{\varepsilon_{t}^{n}|\mathcal{F}_{t-1}\right\}=\exp\left(n \bar{H}_{t}\right)\frac{n!}{1-n^{2}\Delta^{2}} \label{lapmoments}
\end{equation}
for even $n\geq 2$. The right hand side of (\ref{lapmoments}) with $n=1$ additionally gives $E\left\{\left|\varepsilon_{t}\right||\mathcal{F}_{t-1}\right\}$.

\begin{figure}
\includegraphics[width=\linewidth]{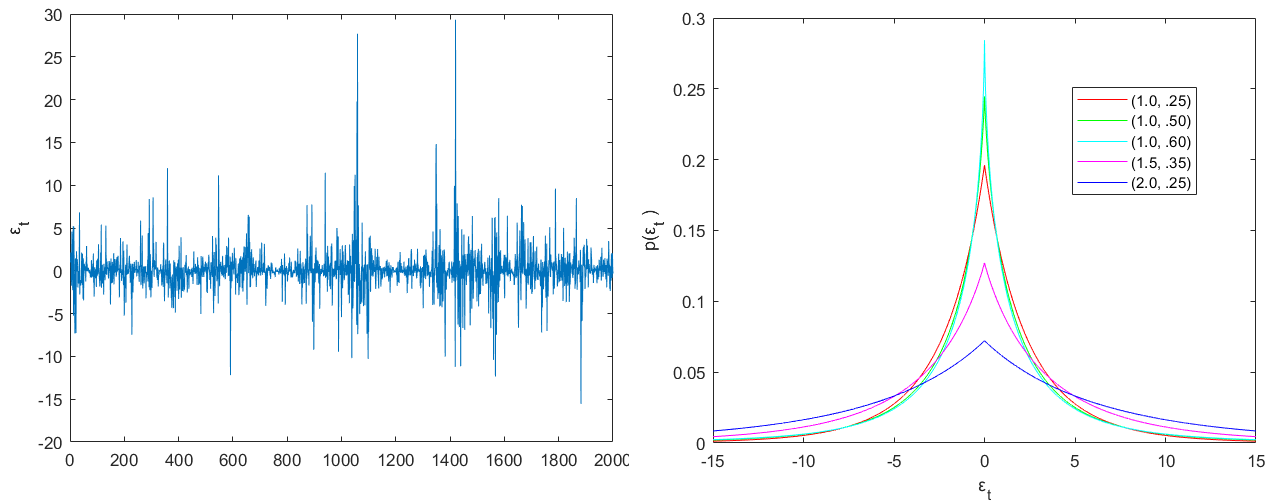}
\caption{Left: A typical realization of the process $\varepsilon_{t}$ defined in (\ref{process})-(\ref{laplace}) with $z_{t}\sim Lap(0,1)$, $H_{t}=.5H_{t-1}+.4H_{t-2}+h_{t}$, and $\Delta=1/4$. Right: Graphs of the probability density functions (\ref{bigproblap}) with $\left(\bar{H}_{t},\Delta\right)$ equal to (1, .25) in Red, (1, .50) in Green, (1, .60) in Cyan, (1.5, .35) in Magenta, and (2, .25) in Blue.}
\label{fig: 4}
\end{figure}

We next note that an equivalent argument to the one in A.3 can be given, using (\ref{loglaplace}) and (\ref{zlap}) with (\ref{productformula}) to give an expression composed of two integrals, each of which can be written very similarly to (\ref{i1}) and (\ref{i2}). Then (\ref{upgammaasympt}) and (\ref{gammaidentity}) can be used in the same way as above to simplify these terms. Adding these two terms gives the result
\begin{eqnarray}
p_{\varepsilon}\left(\varepsilon_{t}|\mathcal{F}_{t-1}\right)=\frac{1}{4 \Delta} \exp\left(-\frac{\bar{H}_{t}}{\Delta}\right)\Gamma \left(1-\frac{1}{\Delta},\frac{\left|\varepsilon_{t}\right|}{e^{\bar{H}_{t}}}\right)\left|\varepsilon_{t} \right|^{1/\Delta-1} \label{bigproblap} \\ \nonumber +\frac{1}{4\Delta}\exp\left(\frac{\bar{H}_{t}}{\Delta}\right) \gamma \left(1+\frac{1}{\Delta},\frac{\left|\varepsilon_{t}\right|}{e^{\bar{H}_{t}}} \right)\left|\varepsilon_{t} \right|^{-1/\Delta-1}
\end{eqnarray}
The identities (\ref{upgamma0}) and (\ref{lowgamma0}) in A.4 can then be used to show that as $\left|\varepsilon_{t}\right| \rightarrow 0$,
\begin{equation}
p_{\varepsilon}\left(\varepsilon_{t}|\mathcal{F}_{t-1}\right)\sim \frac{1}{2 e^{\bar{H}_{t}}}\frac{1}{1-\Delta^{2}}.
\end{equation}
Making the substitution $b=\left|\varepsilon_{t}\right|/e^{\bar{H}_{t}}$ in (\ref{bigproblap}) and applying integration by parts as in A.5 gives the result
\begin{equation}
P\left \{\left|\varepsilon_{t}\right|\geq \Lambda|\mathcal{F}_{t-1}\right\}=\frac{1}{2}\left(\gamma\left(1+\frac{1}{\Delta},\widetilde{\Lambda}_{t}\right)\widetilde{\Lambda}_{t}^{-1/\Delta}-\Gamma\left(1-\frac{1}{\Delta},\widetilde{\Lambda}_{t}\right)\widetilde{\Lambda}_{t}^{1/\Delta}\right)+e^{-\widetilde{\Lambda}_{t}} \label{closedformlap}
\end{equation}
where $\widetilde{\Lambda}_{t}=\Lambda/e^{\bar{H}_{t}}$. Lastly, the argument in A.6 can also be repeated, using (\ref{upgammaseries})-(\ref{ukas}) with the substitutions
$$a_{1}=1-\frac{1}{\Delta}, \hspace{.5cm} a_{2}=1+\frac{1}{\Delta}, \hspace{.5cm} b_{t}=\frac{\left|\varepsilon_{t}\right|}{e^{\bar{H}_{t}}}$$
to write (\ref{bigproblap}) as an asymptotic expansion that is equivalent in form to (\ref{pdf}), with the only change being that the factor $1/\left(4 \Delta \sqrt{2 \pi}e^{\bar{H}_{t}}\right)$ is replaced with $1/\left(4 \Delta e^{\bar{H}_{t}}\right)$. Then re-substituting for $b_{t}$ and $a_{2}$ in the first term gives 
\begin{equation}
\frac{1}{4 \Delta}\exp\left(\frac{\bar{H}_{t}}{\Delta}\right)\Gamma \left(1+\frac{1}{\Delta}\right) \left|\varepsilon_{t} \right|^{-1/\Delta-1},
\label{bigasymptlap}
\end{equation}
which is the limit of (\ref{bigproblap})'s second term as $\left|\varepsilon_{t}\right|\rightarrow \infty$. Multiplying (\ref{bigasymptlap}) by 2 and integrating from $\Lambda$ to $\infty$ will give the first term of our result. Next, fully re-substituting in the second term and remainder term of (\ref{bigproblap})'s asymptotic expansion, applying the integral 
\begin{equation}
\int_{\Lambda}^{\infty}e^{-\varepsilon/e^{\bar{H}_{t}}}\varepsilon^{-(k+1)}d\varepsilon=e^{-k\bar{H}_{t}}\Gamma\left(-k,\frac{\Lambda}{e^{\bar{H}_{t}}}\right), \label{impintlap}
\end{equation}
multiplying by 2, and cancelling factors then gives the full result
\begin{eqnarray}
P\left \{\left|\varepsilon_{t}\right|\geq \Lambda|\mathcal{F}_{t-1}\right\}=\frac{1}{2}\Gamma\left(1+\frac{1}{\Delta}\right)\exp\left(\frac{\bar{H}_{t}}{\Delta}\right)\Lambda^{-1/\Delta} \label{datserieslap}  \\ \nonumber +\frac{1}{2\Delta} \sum_{k=1}^{n-1}\left(-1\right)^{k}\left[\left(\frac{1}{\Delta}\right)_{k}-\left(-\frac{1}{\Delta}\right)_{k} \right]\Gamma\left(-k, \frac{\Lambda}{e^{\bar{H}_{t}}}\right)\\ \nonumber+O\left(\Gamma\left(-n, \frac{\Lambda}{e^{\bar{H}_{t}}}\right)\right).
\end{eqnarray}
Re-applying (\ref{upgammaseries}) to (\ref{datserieslap})'s remainder term with $n=1$ shows that as $\widetilde{\Lambda}_{t} \rightarrow \infty$,
\begin{equation}
P\left \{\left|\varepsilon_{t}\right|\geq \Lambda|\mathcal{F}_{t-1}\right\}=\frac{1}{2}\Gamma\left(1+\frac{1}{\Delta}\right)\widetilde{\Lambda}_{t}^{-1/\Delta} +O\left(\widetilde{\Lambda}_{t}^{-2}e^{-\widetilde{\Lambda}_{t}}\right). \label{pretaillap}
\end{equation}
Then comparing (\ref{pretaillap}) and (\ref{voltail}) gives 
\begin{equation}
P\left \{\left|\varepsilon_{t}\right| \geq \Lambda|\mathcal{F}_{t-1}\right \} \sim \Gamma\left(1+\frac{1}{\Delta}\right) P\left \{\sigma_{t} \geq \Lambda|\mathcal{F}_{t-1} \right \}.  \label{bigtaillap}
\end{equation}
Since $\Gamma(1)=\Gamma(2)=1$, (\ref{bigtaillap})'s second line reduces to $P\left \{\left|\varepsilon_{t}\right| \geq \Lambda|\mathcal{F}_{t-1}\right \} \sim P\left \{\sigma_{t} \geq \Lambda|\mathcal{F}_{t-1} \right \}$
for $\Delta=1$ and as $\Delta \rightarrow \infty$. Several graphs of (\ref{bigproblap}) are given in Fig. \ref{fig: 4}.

\bigskip
\begin{center}
{\large\bf SUPPLEMENTARY MATERIAL}
\end{center}

\begin{description}

\item[R scripts:] R scripts containing code to perform the simulation studies in Section \ref{sec: delta estimation} and Section \ref{sec: lorenz} as well as the empirical application in Section \ref{sec: empirical}. (.R files)

\item[Data sets:] Data sets used in the simulation study in Section \ref{sec: lorenz} and the empirical application in Section \ref{sec: empirical}. (.xlsx files)

\hspace{-1cm}For supplementary materials email gchavez@novocure.com.
\end{description}

\end{document}